\documentclass[twocolumn,numberedappendix,twocolappendix,appendixfloats,numberedaffiliations]{openjournal}

\usepackage{lineno}
\graphicspath{{./}{figures/}}

\usepackage{amsmath}
\usepackage{amssymb}
\usepackage{xspace}
\usepackage{eso-pic}
\usepackage{multirow}
\usepackage{xcolor}
\usepackage{textgreek}
\usepackage[utf8]{inputenc}
\usepackage[T1]{fontenc}
\usepackage[english]{babel}
\usepackage{bigdelim}
\definecolor{linkcolor}{rgb}{0.0,0.3,0.5}
\usepackage{hyperref}
\hypersetup{
    unicode, 
    colorlinks=true,
    linkcolor=linkcolor,
    citecolor=linkcolor,
    filecolor=linkcolor,
    urlcolor=linkcolor,
}
\usepackage[normalem]{ulem}
\usepackage{orcidlink}
\usepackage{soul}

\newcommand{\Gaia}{{\it Gaia}\xspace}

\definecolor{forestgreen}{HTML}{228B22}
\definecolor{urlblue}{HTML}{000000}

\mathchardef\mhyphen="2D

\newlength{\dhatheight}

\newcommand{\code}[1]{\texttt{#1}\xspace}

\newcommand{\unit}[1]{\ensuremath{\mathrm{\,#1}}\xspace}

\newcommand{\degree}{\ensuremath{{}^{\circ}}\xspace}

\newcommand{\asec}{\unit{arcsec}}

\newcommand{\kpc}{\unit{kpc}}

\newcommand{\magn}{\unit{mag}}

\newcommand{\bandvar}[2][]{%
  \ifthenelse{\isempty{#1}}{\var{#2}}{\var{#2\_#1}}%
}

\newcommand{\HEALPix}{\code{HEALPix}}
\newcommand{\healpix}{\HEALPix}

\newcommand{\ugali}{\code{ugali}}

\newcommand{\var}[1]{\ensuremath{\texttt{\MakeUppercase{#1}}}\xspace}
\newcommand{\nside}{\code{nside}}
\newcommand{\RubinSim}{\code{RubinSim}}
\newcommand{\Gala}{\code{Gala}}

\newcommand{\StreamObs}{\code{StreamObs}}
\newcommand{\numpy}{\code{numpy}}
\newcommand{\extend}{\code{EXTENDEDNESS}}

\newcommand{\LSST}{LSST\xspace}
\newcommand{\DES}{DES\xspace}

\newcommand{\AAU}{AAU\xspace}

\newcommand{\Atlas}{Atlas\xspace}
\newcommand{\AliqaUma}{Aliqa Uma\xspace}
\newcommand{\ra}{\ensuremath{\rm RA}\xspace}
\newcommand{\dec}{\ensuremath{\rm Dec}\xspace}
\newcommand{\mtrue}{\ensuremath{m^{\mathrm{true}}}\xspace}
\newcommand{\mlim}{\ensuremath{m^{\mathrm{lim}}}\xspace}

\newcommand{\gband}{g-band\xspace}
\newcommand{\rband}{r-band\xspace}

\providecommand\physrep{\ref@jnl{Phys.~Rep.}}%
\providecommand\apjs{\ref@jnl{ApJS}}%
\providecommand{\jcap}{\ref@jnl{JCAP}}%

\definecolor{darkgreen}{rgb}{0.05,0.3,0.05}
\definecolor{vert}{rgb}{0.1,0.6,0.1}
\definecolor{orange}{rgb}{0.9,0.4,0.2}

\usepackage[dvipsnames]{xcolor}

\shorttitle{LSST Systematics on Stellar Streams}
\shortauthors{P\'{e}lissier et al.\ (LSST Dark Energy Science Collaboration)}

\begin{document}

\title{Impact of LSST systematics on stellar-stream density fluctuations for dark matter}

\author{Matthieu Pélissier\orcidlink{0009-0001-2776-0945}}
\affiliation{Université Grenoble Alpes, CNRS/IN2P3, LPSC, 53 avenue des Martyrs, F-38026 Grenoble, France}
\email{matthieu.pelissier@lpsc.in2p3.fr}

\author{Peter~S.~Ferguson\orcidlink{0000-0001-6957-1627}}
\affiliation{DiRAC Institute and the Department of Astronomy, University of Washington, Seattle, WA, USA}

\author{Alex Drlica-Wagner\orcidlink{0000-0001-8251-933X}}
\affiliation{Department of Astronomy and Astrophysics, University of Chicago, Chicago, IL 60637, USA}
\affiliation{NSF-Simons AI Institute for the Sky (SkAI), 172 E. Chestnut St., Chicago, IL 60611, USA}
\affiliation{Fermi National Accelerator Laboratory, P.O. Box 500, Batavia, IL 60510, USA}
\affiliation{Kavli Institute of Cosmological Physics, University of Chicago, Chicago, IL 60637, USA}

\author{Marine Kuna\orcidlink{0000-0002-3598-2847}}
\affiliation{Université Grenoble Alpes, CNRS/IN2P3, LPSC, 53 avenue des Martyrs, F-38026 Grenoble, France}

\author{David Maurin\orcidlink{0000-0002-5331-0606}}
\affiliation{Université Grenoble Alpes, CNRS/IN2P3, LPSC, 53 avenue des Martyrs, F-38026 Grenoble, France}

\author{Christian Aganze \orcidlink{0000-0003-2094-9128}}
\affiliation{Kavli Institute for Particle Astrophysics \& Cosmology, Stanford University, Stanford, CA 94305, USA}
\affiliation{4 SLAC National Accelerator Laboratory, Menlo Park, CA 94025, USA}

\author{Johann Cohen-Tanugi \orcidlink{0000-0001-9022-4232}}
\affiliation{Université Clermont-Auvergne, CNRS, LPCA, 63000
Clermont-Ferrand, France}

\author{Yao-Yuan Mao \orcidlink{0000-0002-1200-0820}}
\affiliation{Department of Physics and Astronomy, University of Utah, Salt Lake City, UT 84112, USA}

\author{The LSST Dark Energy Science Collaboration}

\makeatletter
\def\@collaboration@present#1#2#3#4{%
 \par
 \begingroup
  \frontmatter@collaboration@above
  \@affilID@def{}%
  \@tempcnta\z@
  \@author@present{}{\ignorespaces#3\unskip}{#4}%
  \par
 \endgroup
 \set@listcomma@list#1%
}%
\makeatother


\begin{abstract}
The Vera C.\ Rubin Observatory's Legacy Survey of Space and Time (LSST) is expected to significantly advance the study of Milky Way stellar streams. In particular, the deep, precise photometry from LSST should greatly increase the statistical sensitivity to density fluctuations in stellar streams, which can be used to probe the small-scale distribution of dark matter. However, current forecasts generally neglect the impact of observational systematics that will be imprinted on stream density measurements.
In this study, we develop a realistic forward-modeling framework to inject stellar streams into LSST-like observations including photometric uncertainties, survey depth variations, background contamination, and imperfect star–galaxy classification. We develop a likelihood-ratio analysis to assess the detectability of gaps in stellar streams in the presence of these observational systematics.
In the presence of realistic survey systematics, we find that after four years of operations, \LSST will be sensitive to density reductions of $\sim50\%$ for gaps with widths of $5\,\deg$ in streams with surface brightness of $\sim33$\,mag\,arcsec$^{-2}$. Relative to the ideal case, this corresponds to a degradation in gap depth sensitivity by a factor of $\sim5$ due to the combined impact of background contamination and observational systematics. 
Assuming a simplified analytical mapping between gap depth and dark matter subhalo properties, these estimates correspond to a minimum detectable subhalo mass of $\sim1\times10^7$\,M$_\odot$. Observational effects shift this accessible mass scale upward by a factor of $\sim16$, with background contamination contributing a factor of $\sim5$ and survey systematics a further factor of $\sim3$, dominated by star-galaxy classification.

These results highlight the need to include realistic observational effects in future stellar stream forecasts and dark matter analyses.

\end{abstract}

\maketitle



\section{Introduction}
\label{sec:intro}

Dark matter (DM) constitutes the dominant component of matter in the Universe and plays a central role in the formation and evolution of cosmic structures \citep[e.g., see review by][]{Bertone_2018}. While the $\Lambda$CDM model successfully describes the large-scale distribution of matter, the microphysical nature of DM remains unknown, and different particle physics scenarios predict distinct behaviors on small scales \citep[e.g.,][]{feng:2010,Bozorgnia_2025}.

In particular, the abundance and properties of low-mass dark matter subhalos are directly linked to the nature of DM and to the process of structure formation on small scales \citep{bullock:2017}. Different DM models predict different subhalo abundances and internal structures. For instance, warm and self-interacting dark matter scenarios are expected to change the abundance and internal structure of low-mass DM subhalos relative to $\Lambda$CDM \citep[e.g.,][]{lovell:2014,huo:2018,amorisco:2022}.

Measurements of the low-mass DM subhalo population therefore provide a powerful probe of DM particle physics. However, below $\sim10^8M_\odot$, the subhalos are expected to contain little or no baryonic matter \citep{benitez-llambay:2020}, preventing direct observations. 
Current observational probes are therefore limited, relying mainly on strong gravitational lensing \citep[e.g.][]{dalal:2002,vegetti:2010,gilman:2020}, pulsar timing \citep{chakrabarti:2026}, or on the dynamical impact of subhalos on stellar systems. In this context, stellar streams provide a unique opportunity to constrain the clustering of DM at small scales \citep{bovy:2017,banik:2021,bonaca:2019,nibauer:2025}.

Stellar streams are elongated stellar structures formed by the tidal disruption of globular clusters or satellite galaxies. As stars are progressively stripped from their progenitor by tidal forces, they spread along the orbit and form coherent structures that can extend for tens of degrees across the sky. In particular, streams originating from globular clusters are dynamically cold systems with low internal velocity dispersion, making them especially sensitive to gravitational perturbations that can distort their morphology and imprint fluctuations along their structure \citep[e.g.,][]{carlberg:2013, newberg:2016, bonaca:2024}.

A variety of mechanisms can perturb stellar streams around the Milky Way, including baryonic structures such as giant molecular clouds \citep{amorisco:2016}, spiral arms \citep{zhou:2026}, a time-dependent Galactic potential \citep{guillaume:2026,arora:2026a}, other globular clusters \citep{ferrone:2025} or dwarfs like the LMC \citep{shipp:2021}, the Galactic bar \citep{pearson:2017, banik:2019}, as well as interactions with dark matter subhalos \citep{ibata:2002}. 
The latter are expected to induce localized signatures such as density fluctuations, gaps, track deviations, or kinematic perturbations along the streams \citep[e.g.,][]{carlberg:2013, erkal:2015}.

Over the last decade, stellar streams have been discovered and characterized in large numbers thanks to wide-field surveys \citep[e.g.,][]{Mateu:2023}. Astrometric and spectroscopic measurements from \Gaia have played a major role by identifying coherent structures through the kinematics of stream member stars \citep[e.g.,][]{price-whelan:2018,malhan:2018}. Massively multiplexed spectroscopic surveys have been critical for performing chemodynamical characterization of streams \citep[e.g.,][]{Li:2022, valluri:2025}. At the same time, deep photometric surveys have proven to be extremely powerful tools for detecting stellar streams through matched-filter techniques and measuring spatial overdensities, especially for faint and distant systems \citep[e.g.,][]{rockosi:2002,shipp:2018}.

The NSF-DOE Vera C.~Rubin Observatory's  Legacy Survey of Space and Time (\LSST) is expected to considerably extend photometric studies of stellar streams due to its unprecedented  depth, area, and photometric precision \citep{ivezic:2019}. The large number of faint stars that will be measured by \LSST should enable the discovery of many new stellar streams and improve the characterization of density fluctuations in already known systems \citep{pearson:2024, romanowsky:2025}. Several studies have explored the sensitivity of LSST stellar stream observations to DM subhalos, either through analytical calculations, numerical simulations, or statistical forecasts \citep{erkal:2015a,drlica-wagner:2019,aganze:2024,lu:2025}. These analyses generally estimate the minimum detectable subhalo mass by studying the detectability of perturbations induced in simulated stellar streams.

However, most analyses make idealized assumptions about the observational data, thereby neglecting realistic survey selection effects and observational systematics. In practice, ground-based photometric surveys such as \LSST have appreciable non-uniformity due to varying environmental conditions. Spatial depth variations will change the number of detected stars across the sky and can imprint artificial density fluctuations along stellar streams unrelated to their dynamical history. Recent studies have highlighted the importance of mitigating such survey-induced fluctuations in large photometric datasets \citep{boone:2026}. Moreover, deep photometric surveys contain a large population of distant unresolved galaxies that increasingly resemble point-like stellar sources at faint magnitudes \citep[e.g.,][]{fadely:2012, Sevilla-Noarbe:2018, Slater:2020}, making them a dominant source of contamination in deep stellar stream observations. Measurements of stellar streams therefore strongly depend on both the survey detection and star-galaxy classification efficiencies, emphasizing the need for accurate classification algorithms to mitigate this background contamination. In recent commissioning data from the Vera C.~Rubin Observatory, nearly $20\%$ of galaxies are misclassified as stars in the magnitude range $24.5 < m_i < 25\magn$, while the fraction of correctly classified stars drops to $50\%$ at $m_i=23.8\magn$ \citep[see Sec.~5.6 of][]{rubindp1:2025}. More generally, these effects emphasize the need for realistic survey-level modeling, including both observational systematics and background contamination, when deriving sensitivity forecasts for stellar streams.

In this work, we develop a realistic forward-modeling framework, \StreamObs,\footnote{\url{https://github.com/LSSTDESC/streamobs}, version 0.1.0} to quantify the impact of \LSST observational effects on the detection of density fluctuations in stellar streams.
Our analysis aims to identify and quantify the dominant observational limitations affecting the detection of density fluctuations in stellar streams and the corresponding sensitivity to DM subhalo interactions. 
We quantify how realistic survey effects degrade sensitivity relative to idealized forecasts and evaluate the expected sensitivity of LSST to gaps in stellar streams. 
Finally, we interpret the LSST stream gap sensitivity in the context of a simple analytical model for stream--subhalo interactions to predict the minimum mass DM subhalo that could induce a measurable gap in a stellar stream measured by LSST.

This paper is organized as follows: we present our technique to simulate \LSST survey systematics in Sec.~\ref{sec:simu}, and the  models to describe (un)perturbed stellar streams and background populations in Sec.~\ref{sec:streammodeling}; we present the experimental protocol, including the sampled stream parameters and survey configurations, in Sec.~\ref{sec:protocol};  
we describe the steps used to generate mock data-like catalogs of stellar stream member stars (and backgrounds) matching the survey properties in Sec.~\ref{sec:genobs};
we discuss the statistical analysis used to assess whether perturbed streams are recognized as such in Sec.~\ref{sec:analysis};
we present our results concerning the impact of the survey systematics on the detectability of perturbations (and hence the minimal mass of the DM subhalos detectable) in stellar streams in Sec.~\ref{sec:results}, and discuss them in Sec.~\ref{sec:discussion}. We then conclude in~Sec.~\ref{sec:conclusion}.

\section{LSST simulations}
\label{sec:simu}

Simulation of realistic survey systematics requires representative \LSST data products. The \LSST Data Challenge 2 (DC2) simulations are adopted for this purpose \citep{desc:2021,collaboration:2025}. DC2 provides a detailed realization of the expected survey performance, including photometric uncertainties, as well as object detection and star-galaxy classification efficiencies. Since DC2 was simulated using a specific 5-year survey duration, we use \RubinSim \citep{yoachim:2025} to estimate depth variations for 1-year and 4-year \LSST data releases, representative of the early survey phase and a later, more homogeneous, stage of the survey, respectively.

These ingredients are then used to propagate true stellar properties into observed quantities and to apply survey selections within a probabilistic framework (see Sec.~\ref{sec:genobs})

\subsection{DC2 simulations}

The DC2 simulations provide both truth and measured catalogs, which are used to characterize the survey response.
The truth catalog contains the intrinsic properties of simulated objects, including noiseless magnitudes, positions, and morphological parameters. The CosmoDC2 galaxy population \citep{korytov:2019} is based on the Outer Rim simulation \citep{heitmann:2019}. The stellar catalog, designed to mimic the Milky Way bulge, disk, and halo populations, relies on \texttt{galfast} \citep{juric:2008}, and is extended to $r>27\,\magn$ by extrapolating SDSS-derived luminosity functions to fainter magnitudes \citep[Sec.~5.3]{desc:2021}.

The measured catalog corresponds to the same objects after passing through the full \LSST image simulation and data reduction pipeline. Realistic images are generated from the truth catalog using \texttt{imSim},\footnote{\url{https://github.com/LSSTDESC/imSim}} which models the dominant sources of observational noise and instrumental response, including sky background, the atmospheric and optical point spread function, and detector-level effects \citep[Sec.~6.1.1]{desc:2021}. The simulated images are then processed with the LSST Science Pipelines to produce object catalogs with realistic noise, detection thresholds, and photometric measurements based on 5 years of observations \citep{desc:2021}.
Objects from the measured catalog are matched to their truth catalog counterparts using the DC2 truth-match catalog \citep[Sec.~3.2]{collaboration:2025}, which follows the positional association method of \citet{sanchez:2020} with adapted parameters.
This matching enables a direct comparison between intrinsic and observed properties, allowing the estimation of photometric uncertainties, detection efficiencies, and classification performance as a function of magnitude and observational conditions.

\subsection{Survey systematics}
\label{sec:survey_systematics}
Realistic survey systematics are modeled at catalog level using the methodology developed to estimate sensitivity to Milky Way ultra-faint satellite galaxies \citep{drlica-wagner:2020a,tsiane:2025,tan:2026}. This approach is directly applicable to stellar streams, which are also similarly analyzed as resolved stellar populations that are affected by photometric uncertainties and selection effects. Survey-related systematics are primarily modeled through spatial variations of the local magnitude limit, while other effects such as point spread function (PSF) variations or foreground stellar density fluctuations are not explicitly included.

\subsubsection{Photometric uncertainties}
\label{sec:err_photometric}

\begin{figure}[t]
\centering
\includegraphics[width=\columnwidth]{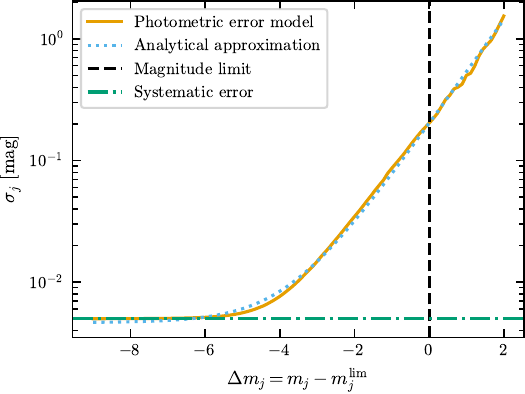}
\caption{Photometric error in $j$-band as a function of the distance to the local magnitude limit estimated from objects in DC2. The model includes a quadratic combination with the systematic errors. The analytical approximation, Eq.~\eqref{eq:error_analytical}, is also presented.
The vertical black line marks the position where the magnitude of the objects reach the survey magnitude limit, corresponding to $\Delta m_j = 0$.}
\label{fig:photometric_errors}
\end{figure}

The photometric error model comes from DC2 as derived by \cite{tsiane:2025} (see their Fig.~2). Uncertainties are parameterized as a function of the distance to the local magnitude limit in the $j$-th band,
\begin{equation}
    \Delta m_j = m_j^{\rm true} - \mlim_j \, ,
    \label{eq:deltamag}
\end{equation}
where $m_j^{\rm true}$ is the true apparent magnitude and $\mlim_j$ the local survey magnitude limit at the object's position. This parametrization captures the degradation of photometric precision near the magnitude limit. To avoid unrealistically small uncertainties for bright sources, we include a minimal photometric uncertainty floor by adding in quadrature a term $\sigma_j^{\mathrm{floor}} = 0.005\,\magn$.
The resulting error model is shown in Fig.~\ref{fig:photometric_errors}. 
It can be approximated with the analytical form
\begin{equation}
    \mathrm{err}(m_j) = a + \exp\!\left(\frac{m_j - b}{c}\right) \, ,
    \label{eq:error_analytical}
\end{equation}
where the parameters $a$, $b$, and $c$ depend on the survey characteristics, in particular the local magnitude limit.

\subsubsection{Stellar detection and classification efficiency}
\label{sec:star_detect_efficiencies}
Selection functions are estimated within the same \mbox{$\sim 3\,\mathrm{deg}^2$} subregion of the DC2 Wide-Fast-Deep footprint that was used in \citet{tsiane:2025}, to ensure consistency and enable direct comparison with their results. Restricting the analysis to a small patch also allows the selection functions to be estimated at effectively fixed magnitude limits, avoiding additional variations induced by survey depth inhomogeneities.
Applying these selection functions to the full survey is justified by their parametrization in terms of the distance to the local magnitude limit. This allows the results obtained in a fixed-depth region to be rescaled across the footprint using the local survey depth, capturing the dominant variations while remaining computationally efficient.

This analysis focuses on point-like sources, which include stellar stream members, foreground stars, and compact galaxies that may be misclassified as stars. In the DC2 catalogs, these objects are identified using the boolean variable \extend \citep{bosch:2018, collaboration:2025}, which encodes a morphological star–galaxy classification based on the consistency between the PSF and CModel magnitudes in the $i$ band. Sources with $\extend = 0$ are classified as point-like, while extended sources exhibit significant differences between the two measurements. Throughout this work, the classification performance obtained with the $\extend$ criterion is referred to as the observed star-galaxy classification (Obs S/G).

As in \citet{tsiane:2025}, the detection and classification efficiencies of stars are estimated in bins of $\Delta m_r$ (see Eq.~\ref{eq:deltamag}):
\begin{equation}
    \mathcal{E}_{\mathrm{detection}}(\Delta m_r^{\rm}) = 
    \frac{N_{\mathrm{detected}}(\Delta m_r)}{N_{\mathrm{true}}(\Delta m_r)} \, ,
    \label{eq:eff_detect}
\end{equation}
\begin{equation}
    \mathcal{E}_{\mathrm{classification}}(\Delta m_r) = 
    \frac{N_{\mathrm{classified}}(\Delta m_r)}{N_{\mathrm{detected}}(\Delta m_r)} \, ,
    \label{eq:eff_class}
\end{equation}
where $N_{\mathrm{true}}$ and $N_{\mathrm{detected}}$ are the number of stars in the DC2 truth and measured catalogs, respectively, and $N_{\mathrm{classified}}$ is the subset of objects correctly classified as point-like ($\extend=0$) and satisfying the quality cut $\sigma_{r} < 0.2$ (i.e., $S/N > 5$) in the $r$ band. This requirement ensures a minimal level of photometric quality, excluding faint sources near the magnitude limit that are more strongly affected by survey systematics and therefore less reliable for analysis. Additional magnitude cuts and selection criteria are applied in the context of this study (see Sec.~\ref{sec:colormagcuts}).

The resulting efficiencies are shown in Fig.~\ref{fig:selection_functions}. They decrease as $\Delta m_r$ gets larger (i.e., the object magnitude approaching the magnitude limit of the survey), reflecting the fact that fainter sources are harder to detect and even more difficult to classify reliably.

\subsubsection{Galaxy misclassification}
\label{sec:galaxy_missclassification}

Background galaxies that are misclassified as stars represent a major source of contamination in stellar stream studies, particularly at faint magnitudes where morphological classification degrades.
In the DC2 catalogs, these contaminants are predominantly compact and/or distant galaxies whose apparent sizes are comparable to the PSF, making them indistinguishable from point-like sources. In contrast, more extended galaxies are less likely to be misclassified due to their broader light profiles.
To ensure a consistent treatment, the analysis is restricted to galaxies with \code{size\_true} $< 0.3\,\mathrm{arcsec}$. This selects only intrinsically compact systems, ensuring that all modeled galaxies lie in the unresolved regime where star–galaxy confusion occurs.

\begin{figure}[t]
\centering
\includegraphics[width=\columnwidth]{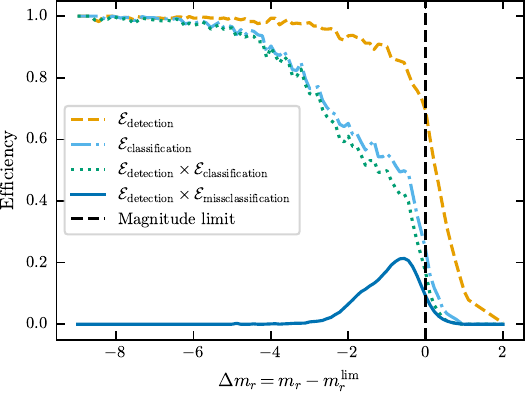}
\caption{Various survey selection efficiencies estimated from the DC2 catalogs as a function of $\Delta m_r$, defined in Eq.~\eqref{eq:deltamag}.
The detection efficiency for point-like sources (orange dashed line), the classification efficiency for stars (blue dashed-dotted line), their product (green dotted line), and the combined detection and misclassification efficiency for compact galaxies (blue solid line) are shown.
The vertical black line marks the position where the magnitude of the objects reach the survey magnitude limit, corresponding to $\Delta m_r = 0$.}
\label{fig:selection_functions}
\end{figure}

The misclassification efficiency in a $\Delta m_r$ bin is
\begin{equation}
    \mathcal{E}_{\mathrm{misclassification}}(\Delta m_r) = 
    \frac{N_{\mathrm{misclassified}}(\Delta m_r)}{N_{\mathrm{detected, galaxies}}(\Delta m_r)} \, ,
    \label{eq:eff_missclass}
\end{equation}
where $N_{\mathrm{detected, galaxies}}$ is the number of true galaxies detected in the measured catalog, and $N_{\mathrm{misclassified}}$ the number of detected galaxies that are classified as point-like ($\extend=0$) and pass the cut $\sigma_{r} < 0.2$. The combination of detection and misclassification efficiencies is also displayed in Fig.~\ref{fig:selection_functions}. It increases near the magnitude limit, where morphological classification reaches its limits. Indeed, it rises to a maximum just before reaching the magnitude limit and then decreases with increasing $\Delta m_r$, due to the additional signal-to-noise cut $\sigma_{r} < 0.2$ (i.e., $S/N > 5$) applied to the sample.
Although the misclassification efficiency for galaxies is lower than the classification efficiency for stars, galaxies are far more numerous, making them a dominant source of background contamination (see Sec.~\ref{sec:colormagcuts}).

\subsubsection{Depth maps}
\label{sec:depth_maps}
The local magnitude limit $\mlim_j$ in any band $j$ provides maps of the $S/N=5$ limiting magnitude across the \LSST footprint for different data releases. These maps incorporate the survey cadence, observing conditions, and coaddition strategy. They are generated with \RubinSim \citep{yoachim:2025}, using the \texttt{baseline\_v5.0.0} cadence simulation. At each sky position, the depth results from coadding simulated visits up to the relevant data-release epoch, accounting for sky brightness, seeing, and exposure time. The maps capture survey cadence, observing conditions, and coaddition strategy, but exclude Galactic dust extinction, which is applied independently at the catalog level (Sec.~\ref{subsubsec:dust}). The depth maps are provided as \texttt{HEALPix} maps and distributed as ancillary data through the \texttt{StreamObs} repository.

In this study, we focus on making projections for several early data releases from \LSST. These early releases are expected to be shallower than the DC2 simulations that were used to derive the survey selection functions and photometric uncertainties. This allows us to reliably study survey configurations without extrapolating beyond the model regime explored in DC2. In particular, we consider releases corresponding to the first year of \LSST\ observations, representative of the initial survey products, and to the first four years, when the survey coverage becomes more homogeneous.

\begin{figure*}[t]
\centering
\includegraphics[width=\textwidth]{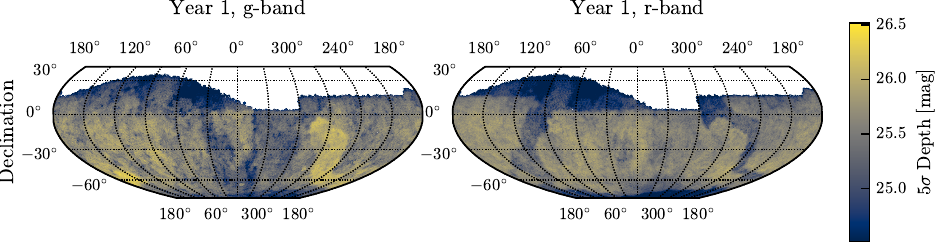}\\
\includegraphics[width=\textwidth]{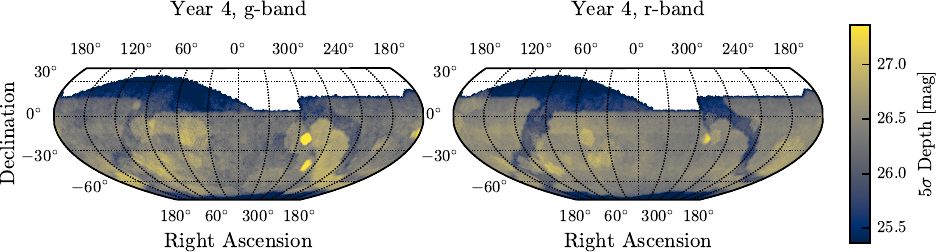}
\caption{Maps of the $S/N=5$ limiting magnitude predicted by \RubinSim for the
\LSST survey. Top panels correspond to 1 year of observation and bottom panels
to 4 years. Left and right columns show the \gband and \rband, respectively.}
\label{fig:depth_maps}
\end{figure*}

For illustration, we show in Fig.~\ref{fig:depth_maps} the depth maps in the \gband (left panels) and \rband (right panels) after 1 year (Year 1) on top panels and 4 years (Year 4) on the bottom panels. The deepest regions correspond to some Target of Opportunity observations and deep-drilling fields, while the shallowest lie along the North Ecliptic Spur and the Galactic plane.  Those maps are the foundation of the survey systematics modeling, as they determine the spatial variations of photometric uncertainties and selection functions across the survey footprint in our model.

To also probe an idealized case in which there are no systematic variations in depth, we also constructed uniform depth maps with the same footprint as each release, assigning a constant magnitude limit that is equal to the median depth within the mask of interest (see Sec.~\ref{subsec:position}).

\section{Stellar stream modeling}
\label{sec:streammodeling}
Throughout this section, the stream, gap, and background are modeled using the configuration most favorable to detection: a constant distance (Sec.~\ref{sec:stream_isochrones}), a uniform baseline density perturbed by a three-parameter box-car gap (Sec.~\ref{sec:gap-model}), and a spatially uniform contaminant background (Sec.~\ref{sec:bkdg}); the impact of these simplifications is quantified in Sec.~\ref{sec:limitations}. These choices are deliberate: the goal of this study is to isolate and quantify the response of the survey itself to density fluctuations, rather than to capture the full dynamical complexity of stream and background morphology. Working with the fewest stream and background-free parameters keeps the hypothesis test of Sec.~\ref{sec:analysis} tractable over the thousands of realizations analyzed in Sec.~\ref{sec:results}, while placing the analysis in the regime where a genuine perturbation is easiest to detect, with the detection efficiency driven solely by survey systematics instead of stream dynamics.

\subsection{Isochrone, distance and width}
\label{sec:stream_isochrones}

Dynamically cold stellar streams are generally well described by a single old and metal-poor stellar population. As a representative template for such systems, we adopt the \Atlas–\AliqaUma (\AAU) stream \citep{koposov:2014, shipp:2018, Li:2021}, a promising stream located within the \LSST footprint. This choice is also motivated by the fact that stellar streams on wide orbits with large pericentric distances are expected to be less affected by baryonic perturbations and are therefore particularly well suited for probing DM subhalo interactions \citep{hilmi:2024, nguyen:2025}. Furthermore, \AAU\ exhibits properties typical of the low-surface-brightness streams that \LSST is expected to detect \citep{drlica-wagner:2019}.

The parameters of the stellar population for the simulated stream are fixed to those inferred for \AAU in \cite{shipp:2018}. The stream is modeled with an isochrone of metallicity $[\mathrm{Fe/H}] = -1.35$ and age $11\,\mathrm{Gyr}$ using the PARSEC model from \cite{marigo:2017} with the standard LSST bandpasses (R1.9) integrated into the Padova CMD interface\footnote{\url{http://stev.oapd.inaf.it/cgi-bin/cmd}}. The stellar population is sampled using \ugali\footnote{\url{https://github.com/DarkEnergySurvey/ugali}} \citep{Bechtol:2015, drlica-wagner:2020a}, a stellar population modeling and satellite search package.
We make the simplifying assumption that the stream lies at a constant distance from the observer, corresponding to a distance modulus of $m-M = 16.8\,\magn$ ($22.9\,\kpc$), without including any distance gradient along the stream. The stream is assigned a Gaussian width of $0.25^\circ$ (corresponding to $\sim 95\,\mathrm{pc}$), consistent with observations \citep{shipp:2018}.

\subsection{Baseline vs perturbed stream modeling}
\label{sec:gap-model}
To quantify the sensitivity to density fluctuations, we adopt a simple parametric description of the stellar density along the stream-aligned coordinate $\phi_1$, where $(\phi_1,\phi_2)$ denote the on-sky coordinates respectively parallel and perpendicular to the stream track. While full dynamical simulations provide a more realistic modeling of subhalo-stream interactions, they are computationally expensive and less suited for systematically exploring detection limits over a wide parameter space. The parametric approach, instead, allows for controlled injections of fluctuations and a direct quantitative assessment of the fluctuation detection efficiency. Moreover, the analysis pipeline is designed to be fully compatible with inputs from dynamical simulations, which can be incorporated in future studies to refine these results.

\subsubsection{Gap model distribution function}
\label{sec:gap_model}
For simplicity, we write the on-sky probability density function of the stars (in the stream along $\phi_1$) as the product of the baseline stream (Base) and a modulation function (Mod), up to a normalization constant,
\begin{equation}
    P(\phi_1) \propto {\rm Base}(\phi_1) \times {\rm Mod}(\phi_1)\,,
    \label{eq:pdf}
\end{equation}
with the baseline density given by a top-hat profile
\begin{equation}
    {\rm Base}(\phi_1) = 
    \begin{cases}
        1 & \text{if } \phi_1 \in [-l/2,\, l/2] \\
        0 & \text{otherwise}
    \end{cases} \,,
\end{equation}
with $l$ representing the stream length in degrees. 

The current work focuses primarily on a simple single (on-sky) gap model, namely
\begin{equation}
    {\rm Mod}[A\,,x_{\mathrm{gap}},\, w] =
    \begin{cases}
        1 & \text{if~} |\phi_1-x_{\rm gap}|> \frac{w}{2} \\
        1 - A & \text{otherwise}
    \end{cases},
    \label{eq:gapmodel}
\end{equation}
where $A$ is the amplitude of the gap, $x_{\rm gap}$ is its position, and $w$ is its width.
Although DM subhalo impacts result in more complex density structures \citep{erkal:2015}, the gap model captures the leading feature of a localized density deficit. It provides a practical template for data fitting, as illustrated in \citet{bonaca:2019}. We further restrict the analysis to a single visible gap, consistent with expectations for stellar stream--subhalo interactions \citep{erkal:2016}. Hence, the gap model is simple enough to allow a study of its detectability against the survey systematics, while the width and amplitude of the gap can be linked to realistic DM subhalo perturbations (see Appendix~\ref{app:subhalosproba}). 
Figure~\ref{fig:modelexample} shows the linear density in this model for several values of the gap amplitude $A$.

\subsubsection{Fixed $N_\star$ or fixed baseline density?}
From the probability density function $P(\phi_1)$ defined in Eq.~\eqref{eq:pdf}, the expected linear density along the stream, $ \lambda_n$, can be computed in discrete bins of width $b_n$ of indices $n$,
\begin{equation}
    \lambda_n = N_\star \int_{b_n} \frac{P(\phi_1)}{b_n}\,\mathrm{d}\phi_1 \, ,
    \label{eq:lindens_model}
\end{equation}
where $N_\star$ denotes the total number of stream stars. The bin size is chosen to $b_n = 0.5\degree$, to reduce shot noise in the measurements.

\begin{figure}[t]
\centering
\includegraphics[width=\columnwidth]{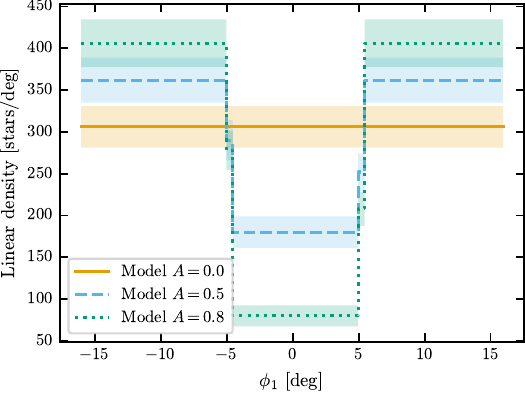}
\caption{Illustration of the gap model linear density for a stream length $l=30\degree$. The three curves show different gap amplitudes $A$ at a fixed width $w=5\degree$ and position $x_{\rm gap}=0$. Each model is normalized, see Eq.~\eqref{eq:lindens_model}, so that the total number of stars ($N_\star =10^{4}$) remains the same for all gaps amplitudes. The color-bands correspond to the Poisson uncertainties.}
\label{fig:modelexample}
\end{figure}

We choose to study density fluctuations at a fixed total number of stars $N_\star$, by comparing realizations of a uniform stream ($A=0$) to perturbed streams with $A>0$. In all cases, the total number of stars is conserved. As a result, varying the fluctuation amplitude does not correspond to removing stars, but rather to redistributing them along the stream. In practice, an increase in $A$ produces both a depletion within the gap and a corresponding excess in the surrounding regions, ensuring number conservation. This behavior is illustrated in Fig.~\ref{fig:modelexample}, where the formation of a gap is accompanied by an overdensity outside the gap.

This choice is motivated by the expected physical origin of gaps induced by DM subhalos, where stars are displaced along the stream rather than removed \citep{erkal:2015}.
It also ensures a consistent definition of the detection problem, interpretable either statistically, as a deviation from uniform density at fixed star count, or photometrically, as a deviation from uniform surface brightness at fixed total flux.
In this regime, redistribution and removal lead to similar observational signatures for small-amplitude perturbations.
For example, a gap with depth $A = 0.4$ and width $w = 3\degree$ induces a change of $3.8\%$ in the surface brightness outside the gap, making the two descriptions effectively degenerate at the level of our analysis.

\subsubsection{Normalizing $N_\star$ to observed streams}
\label{sec:Nstar_norm}
To calibrate the intrinsic number of stars injected in our stream realizations, we anchor the simulations to the observed properties of \AAU. We generate mock observations with \StreamObs assuming a \DES Y3-like survey with $5\sigma$ magnitude limits $m_g^{\mathrm{lim}} = 25.05$ and $m_r^{\mathrm{lim}} = 24.75$ \citep{sevilla-noarbe:2020}, together with color--magnitude selections consistent with \cite{shipp:2018}. Different values of the intrinsic number of stream stars $N_\star^{\mathrm{true}}$ are tested, and the corresponding observed number of stars after survey selection is compared to the number measured for \AAU\ in DES data \citep{shipp:2018}. This procedure leads to a reference value of $N^{\mathrm{true},\,\mathrm{AAU}}_\star \simeq 4\times10^4$ stars for an \AAU-like stream. In the rest of this work, we nevertheless explore a broader range of intrinsic stellar populations in order to study the dependence of our results on stream surface brightness.

The true number of stars can be converted into a total stellar mass by estimating the mean stellar mass from the adopted isochrone. This is obtained by integrating the stellar mass function along the isochrone to derive $\langle M \rangle$, and computing $M_{\mathrm{true}} = N^{\mathrm{true}}_\star \times \langle M \rangle$. Applying this procedure yields an average mass $\langle M \rangle^{\mathrm{AAU}} \simeq 2.4\times 10^{-1}M_\odot$, and a total stellar mass of order $M_{\mathrm{true}}^{\mathrm{AAU}} \sim 10^{4}\,M_{\odot}$, consistent with the stellar mass currently present in the tidal debris of such globular cluster streams \citep{Li:2021,hilmi:2024}.

This stellar population can also be expressed in terms of an average surface brightness. The latter is defined as the integrated apparent magnitude of all stream stars, normalized by the on-sky stream area between $\pm 1 \sigma$ in $\phi_2$. The integration is performed over stars brighter than $m_r < 25$ in $r$-band, but does not change for different values since the total luminosity is dominated by the brightest stars. For the \AAU configuration used in this work, $N^{\mathrm{true}}_\star =10^{4}$ corresponds to a mean surface brightness of $\mu_r = 32.90\,\mathrm{\magn\,\asec^{-2}}$ in $r$-band, which agrees well with the value that \citet{shipp:2018} quote for \Atlas ($\mu_V = 33.0\,\mathrm{\magn\,\asec^{-2}}$).

\subsection{Contaminants}
\label{sec:bkdg}
Stellar streams span large areas on the sky and are heavily contaminated by Milky Way foreground stars and unresolved background galaxies. Isochrone-based color--magnitude selection can reduce contamination but never fully remove it. 
Therefore, a realistic modeling of both stellar and galactic contaminant populations is required.
We assume both populations to be spatially uniform across the sky.
This approximation is sufficient for the present study---which focuses on survey-induced systematics at high Galactic latitudes. Indeed, it provides a controlled baseline corresponding to an idealized homogeneous contaminant population.

The intrinsic magnitude distribution of the contaminants is derived from DC2, within a $1\degree$ radius region centered on the patch used in \cite{tsiane:2025}. We apply a true magnitude cut $m_g < 28$ to define the underlying catalog. This is a conservative choice with respect to the expected \LSST depth after 1 and 4 years of observations. For galaxies, we further restrict the sample to compact objects with $\code{size\_true} < 0.3\,\mathrm{arcsec}$. This selects stars and galaxies that could plausibly be misclassified (as point-like) sources.

From this truth-level catalog, we derive surface densities for each population, obtaining $\sim 10^{6}\,\mathrm{deg}^{-2}$ galaxies and $\sim 10^{4}\,\mathrm{deg}^{-2}$ stars. These values correspond to intrinsic densities, prior to any survey selection effects.
For a given simulated stream, we define an analysis window corresponding to the angular extent along and across the stream (discussed in Sec.~\ref{sec:stream_isochrones} and \ref{subsec:position}, respectively). We sample the corresponding number of background objects from these surface densities. This ensures that each realization contains a statistically consistent background population.
Finally, the multi-band magnitudes of the sampled objects are assigned by bootstrapping from the DC2 truth catalog distributions, preserving realistic stellar and galactic color--magnitude properties at the input (truth) level. This choice is motivated by the fact that the available truth catalog may contain fewer objects than required to populate the simulated background over the full stream region; resampling with replacement therefore ensures sufficient statistics while maintaining the empirical distributions.
In this way, we generate uniform contaminant populations with realistic distributions in color--magnitude space.

\section{Stellar streams in realistic catalogs}
\label{sec:genobs}
To quantify the \LSST sensitivity to density fluctuations in stellar streams, mock data sets are generated with realistic survey systematics using the public software \StreamObs \citep{streamobs}. \StreamObs is a forward-modeling package designed to transform intrinsic stellar populations into observed data-like catalogs for photometric surveys. It propagates true stellar properties through
survey-specific effects, including depth variations, photometric uncertainties, selection functions, and classification performance. The package is under active development, with planned extensions including support for additional surveys and more advanced modeling of observational systematics and dynamical stream inputs.

We summarize below the main steps used to generate mock data-like catalogs. 
The generation stage (Sec.~\ref{sec:mock_generation}) constructs stellar streams and background populations at a given position on the sky, and assigns observed quantities such as magnitudes and photometric uncertainties. These incorporate both survey depth variations and dust extinction effects, ensuring a realistic description of the observed data.
In the reconstruction stage (Sec.~\ref{sec:moc_selection}), objects are probabilistically classified as detected or not based on these properties. The resulting catalog is then processed using selection criteria designed to enhance the contrast between stream stars and contaminants (Sec.~\ref{sec:SN_optimization}). This mimics a realistic analysis pipeline, optimizing the signal-to-noise ratio of the stream and enabling the subsequent gap detection analysis (Sec.~\ref{sec:analysis}).

\subsection{Generation step}
\label{sec:mock_generation}
Synthetic source injection is commonly used to model survey selection effects in a fully realistic way, by injecting artificial sources into imaging data and processing them through the full detection and measurement pipeline (e.g., \citealt{everett:2022, boone:2026}). This approach naturally captures complex observational effects, including crowding, blending, and instrumental systematics, and therefore provides the most accurate description of survey performance.

However, this method is computationally expensive, especially when large numbers of realizations are required. In the context of stellar stream studies, thousands of mock streams covering $>$10,000 deg$^2$ must be generated and analyzed, either in a parametric framework as done here, or using dynamical simulations for DM constraints \citep{banik:2021, ma:2025,nguyen:2025}. Running synthetic source injection over such large sky areas is computationally infeasible.

Instead, we adopt an approximate approach in which selection functions and photometric uncertainties are modeled as functions of the distance to the local magnitude limit. This provides a computationally efficient alternative that retains the dominant dependence on survey depth, while enabling the generation and analysis of thousands of realizations. This trade-off allows us to explore the impact of survey systematics at the scale required for stellar stream studies, while remaining tractable.

\subsubsection{Stellar stream position}
\label{subsec:position}

Survey systematics are applied by embedding each mock stream at a specific location within the \LSST footprint, defined as the sky area covered by all photometric bands of the survey. This footprint shape evolves only marginally between data releases, as illustrated in Fig.~\ref{fig:depth_maps}.

As our goal is to assess the impact of spatially varying
survey systematics on fluctuation detectability, each stream realization is
placed at a random position\footnote{The intrinsic stream and background positions, defined in the stream-aligned coordinate system $(\phi_1, \phi_2)$, are transformed to equatorial coordinates $(\ra, \dec)$ using the \Gala package 
\citep{price-whelan:2017,price-whelan:2025}.} (drawn uniformly) within the accessible
\LSST footprint. This procedure samples a broad range of observing conditions,
including depth variations, photometric uncertainties, and star-galaxy
classification performance.

Moreover, we avoid stream locations that would intersect regions of high dust extinction, in order to isolate survey systematics from strong reddening effects. Dust extinction is modeled using the $E(B-V)$ maps from \cite{schlegel:1998}, provided as \healpix\ maps with $\nside=512$. We exclude regions with $E(B-V) > 0.2$, which are predominantly located near the Galactic plane.

\begin{figure}[t]
\centering
\includegraphics[width=\columnwidth]{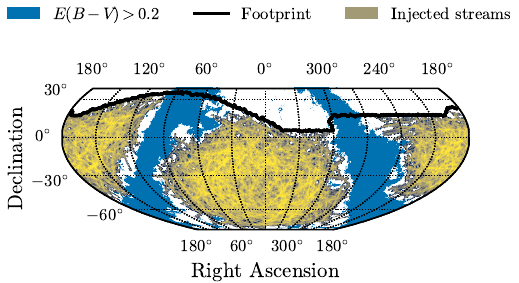}
\caption{Geometric mask used to define the allowed locations of simulated streams. 
The yellow color map shows the locations of 1,500 stream realizations injected within the \LSST footprint (black line), excluding high extinction regions (i.e., $E(B-V) > 0.2$; blue).}
\label{fig:mask}
\end{figure}

The final geometric mask over which we simulate stellar streams is shown in Fig.~\ref{fig:mask}. This mask combines the survey footprint with the low-extinction selection (blue region). In practice, each realization is required to have at least $99\%$ of its stars within the unmasked area, ensuring that measured density fluctuations are not driven by edge effects. As a result, streams are less frequently placed near the boundaries of the mask, since they must lie almost entirely within the allowed region. The yellow color scale represents the spatial distribution of stars from $1500$ stream realizations falling within the mask.

\subsubsection{Observed magnitudes with dust extinction}
\label{subsubsec:dust}

We generate observed magnitudes from the positions and true magnitudes of the simulated objects. For each point-like object of the truth catalog (stream stars, foreground stars, and compact galaxies), simulated observed magnitudes are generated taking into account the dust extinction and photometric uncertainties at the location of each object (see Sec.~\ref{sec:survey_systematics}). The extincted true magnitude is first defined as
\begin{equation}
m_j^{\rm true-uncorr} = m_j^{\rm true} + A_j(\mathrm{RA}, \mathrm{Dec}) ,
\end{equation}
where $A_j(\ra, \dec)=R_j\times E_{B-V}(\ra,\dec)$ is the extinction in the $j$-th band. Here, $j$ denotes the photometric band, $\mtrue_j$ the true magnitude, $E_{B-V}(\ra, \dec)$ the dust reddening at the object position. The coefficients $R_j$ are computed using \RubinSim, yielding $R_g \simeq 3.66$ and $R_r \simeq 2.70$, assuming a fixed $R_V = 3.1$ uniform across the footprint.

The observed quantities are estimated from the distance to the survey magnitude limit at the object’s location. Since the patch used to estimate survey properties had negligible extinction, Eq.~\eqref{eq:deltamag} is generalized to
\begin{equation}
\Delta m_j = m_j^{\rm true-uncorr} - \mlim_j(\ra,\dec)
\end{equation}
where $\mlim_j(\ra, \dec)$ is the local survey magnitude limit. 

The observed flux is obtained by drawing from a Gaussian distribution centered on the flux corresponding to $m_j^{\rm true-uncorr}$, with standard deviation given by this photometric uncertainty $\sigma_j(\Delta m_j)$ estimated from DC2 simulations (see Sec.~\ref{sec:err_photometric} and Fig.~\ref{fig:photometric_errors}). This sampled flux
is converted back into magnitude, yielding the observed (but uncorrected) magnitude $m_j^{\rm obs-uncorr}$ in the $j$-th band.
Finally, to mimic a realistic stellar stream analysis, these magnitudes are corrected for dust extinction:
\begin{equation}
 m_j = m_j^{\rm obs-uncorr}- A_j(\ra, \dec) \, .
\end{equation}
This procedure preserves the degradation of photometric uncertainties induced by dust absorption, while removing the global shift in the magnitude distribution.

We stress that the corrections are performed using the same dust map as that used to perturb the observed quantities, corresponding to an idealized scenario in which the extinction is perfectly known. In practice, uncertainties in dust maps, as well as spatial variations in $R_V$, would introduce additional systematics. The present pipeline could be used to quantify their impact, but we believe that is a subdominant correction compared to the survey systematics at the high Galactic latitudes studied here.

\subsection{Selection step}
\label{sec:moc_selection}
\subsubsection{Survey selection functions}

To determine whether a given object is detected and classified as a point-like source, we apply the efficiencies defined in Eqs.~\eqref{eq:eff_detect} and \eqref{eq:eff_class} within a probabilistic framework. For each object, a random number is drawn from a uniform distribution in the range $[0,1]$. If this value is smaller than the combined detection and classification efficiency evaluated at the object's $\Delta m_r$, the object is considered detected and classified as a point-like source. 

For galaxies, the misclassification efficiency, Eq.~\eqref{eq:eff_missclass}, is applied in the same way to determine if they are erroneously classified as a star. In the case of an idealized perfect star-galaxy classification, only the detection is applied to stars, and no galaxy contaminants leak into the star sample.

Since the selection functions are derived using the $r$-band as the reference band, no additional magnitude-dependent efficiency is applied in the $g$ band to avoid introducing redundant and strongly correlated selection effects between filters. However, because the color-magnitude selection relies on both bands, we impose an independent requirement $\sigma_g < 0.2$ (corresponding to $S/N > 5$) in the $g$ band to ensure that only well-measured point-like sources are used in the analysis.
This cut is applied at the simulated observed catalog level. Objects satisfying these probabilistic criteria and selection cuts are referred to as \textit{observed} throughout this work.

For a stream generated with properties similar to \AAU\ (see Sec.~\ref{sec:streammodeling}), a realistic survey selection accounting for both detection and star-galaxy classification with a $5\sigma$ magnitude limit of $26\magn$ in both $r$ and $g$ bands yields only $\sim12.4\%$ of the intrinsic stream stars observed, while $\sim28.7\%$ (resp. $17.6\%$) of the stars satisfy $m_r^{\mathrm{true}}<26$ (resp. $m_g^{\mathrm{true}}<26$). The matched filter selection would then select 91\% of those observed stars for such a stream (see Sec.~\ref{sec:colormagcuts}), showing that the dominant loss of statistics occurs at the detection and classification stage. This highlights the importance of modeling realistic survey selection functions, and shows that simply selecting stars down to the nominal $5\sigma$ magnitude limit provides an overly idealized description of the observed stream population.

\subsubsection{Stellar stream selection functions}
\label{sec:colormagcuts}
Several selections in color and magnitude are used to enhance the contrast between stream members and contaminants.

\paragraph{Isochrone Matched Filter}
Following \cite{shipp:2018}, a matched filter is constructed around the stream isochrone that is determined using the true distance modulus, age and metallicity of the simulated stream (see Sec.~\ref{sec:streammodeling}). The matched filter shape around that isochrone is parameterized as follows:

\begin{equation}
\begin{aligned}
    (m_g - m_r)^{\mathrm{iso}} 
    &+ E \times \mathrm{err}\!\left(m_g^{\mathrm{iso}} + \frac{\Delta\mu}{2}\right) - C \\
    &< (m_g - m_r) < \\
    (m_g - m_r)^{\mathrm{iso}} 
    &+ E \times \mathrm{err}\!\left(m_g^{\mathrm{iso}} - \frac{\Delta\mu}{2}\right) + C \, ,
\end{aligned}
\label{eq:matchfilter}
\end{equation}
where $(m_g - m_r)^{\mathrm{iso}}$ is the color predicted by the isochrone at a given magnitude, $E$ is the error broadening factor, $C$ the intrinsic color width, and $\Delta \mu$ the distance modulus tolerance.
The parameters are fixed to $E = 2$ and $\Delta\mu = 0.5$, following \cite{shipp:2018}. The color width is chosen to be symmetric with $C = 0.05$. The red giant branch is excluded from the filter, as the small number of RGB stream members provides insufficient signal-to-noise to overcome the heavy foreground contamination.

As in \cite{shipp:2018}, the photometric uncertainty is parametrized analytically as defined in Eq.~\eqref{eq:error_analytical}. The constants of this model are fitted to the \StreamObs error model (see Fig.~\ref{fig:photometric_errors}) with the limiting magnitude $m_g^{\mathrm{lim}}$ set to the $95$-th percentile of the local depth distribution of the selected objects. This choice ensures that the error model reflects the majority of the observed population and adapts to the stream location within the survey.

\begin{figure}[t]
\centering
\includegraphics[width=\columnwidth]{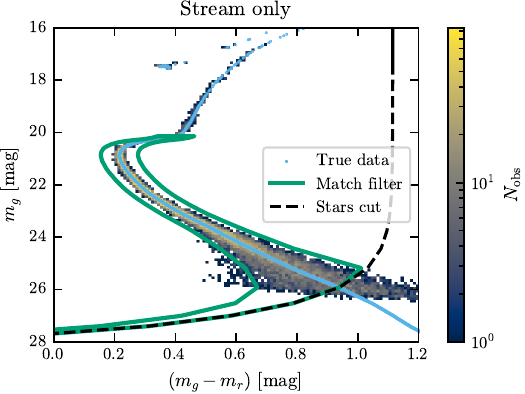}\\
\includegraphics[width=\columnwidth]{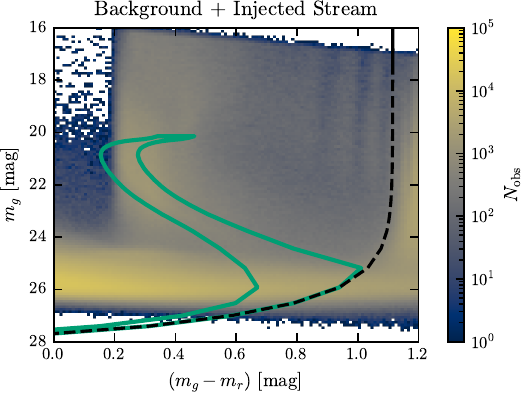}
\caption{Color-magnitude diagram of stream stars alone (top) and combined with stellar and galactic backgrounds (bottom), as generated by \StreamObs for a uniform 4-year \LSST survey and stream with $N^{\mathrm{true}}_\star=10^5$ stars. The green solid line shows the matched filter, while the dashed line indicates the additional component used to suppress the reddest background stars.}
\label{fig:matchfilter}
\end{figure}

Figure~\ref{fig:matchfilter} illustrates the structure of the matched filter and selected star and galaxies samples in color--magnitude space, before and after the impact of \LSST observations.
In the top panel, the filter closely follows the isochrone sequence defined by the true data (green selection region, centered on the isochrone traced by the blue points).
The color--magnitude distribution of observed stars presents an increasing spreading with the magnitude. This is due to fainter stars getting closer to their local magnitude limit, leading to larger errors. It also presents a cut-off at a magnitude around $m_g \sim 26$, introduced by the selection functions near the magnitude limit. The cut in the bright end is due to the survey saturation at $m_g \sim 16$. 
In the bottom panel, the background populations occupy a broader region of the color-magnitude diagram. Compact background galaxies can be seen to extend horizontally across color space at $m_g \sim 26$. At brighter magnitudes, the contamination is primarily due to Milky Way main-sequence stars. 

\paragraph{Mitigation of red stars contamination}

Another noticeable contaminant comes from the plume of M-dwarfs that can be seen at $g - r \sim 1.2$.
At faint magnitudes, the M-dwarf plume broadens due to larger color uncertainties and can in fact dominate the background in the case of perfect star-galaxy classification. For the \AAU stream at $\sim23\,\mathrm{kpc}$, no genuine stream members are expected in this color region; the red plume is therefore purely a foreground population.
Hence, an additional empirical cut is applied,
\begin{equation}
    (m_g - m_r) < 1.12 - \mathrm{err}(m_g) \, ,
    \label{eq:addcut}
\end{equation}
which suppresses the reddest stellar contaminants, while only removing a small portion of stellar stream stars (black dashed line on Fig.~\ref{fig:matchfilter}).

Overall, in the example illustrated in Fig.~\ref{fig:matchfilter}, with an additional spatial cut in $|\phi_2| < 1.4\,\sigma = 0.35^\circ$, the matched-filter selection together with the red-star mitigation cut yields a sample composed of only $\sim16\%$ stream stars, compared to $\sim33\%$ foreground stars and $\sim51\%$ galaxies, highlighting the dominant contamination at faint magnitudes.

\begin{figure*}[t]
\centering
\includegraphics[width=\textwidth]{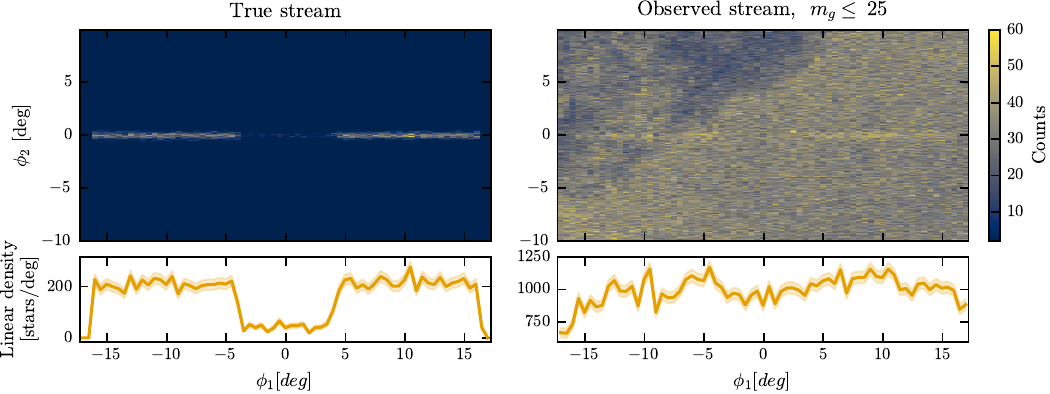}
\vspace{0.5cm}
\includegraphics[width=\textwidth]{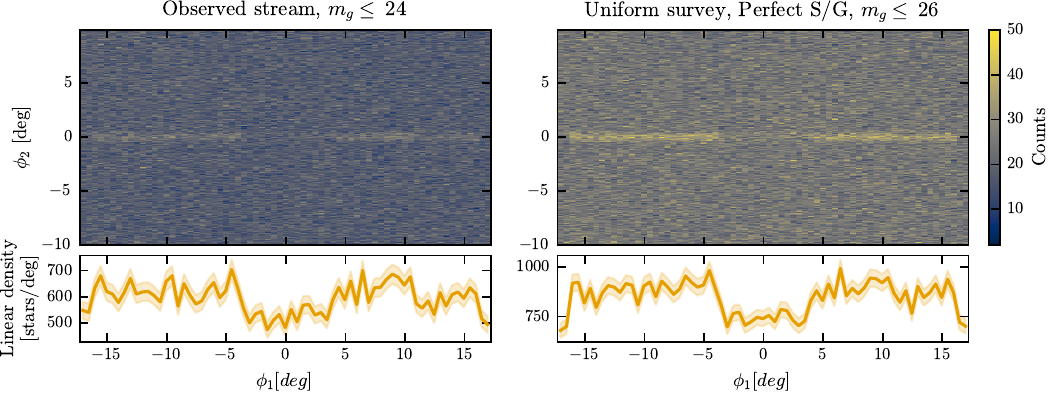}
\caption{
Four observational configurations showing the density of the {\em observed} two-dimensional stellar stream density (top) and the projected (bottom) `on-stream' selection (see text). 
The upper set of panels shows the stream before injection (left) and after forward modeling into the \LSST Year~4 survey using \StreamObs (right). 
The lower set compares a realistic \LSST Year~4 survey (left) with a uniform survey assuming perfect star-galaxy classification (right). 
The one-dimensional density is computed within $|\phi_2| < 1.4\,\sigma = 0.35^\circ$, with the color-bands corresponding to the Poisson uncertainties.
A shallow magnitude cut is applied to enhance the signal-to-noise ratio of the observed stream.
In the realistic case, an additional bright cut is used to mitigate survey systematics.
}
\label{fig:2Ddens}
\end{figure*}

\subsection{Mock observations and systematics mitigation strategy}
\label{sec:SN_optimization}
It is instructive to see how the forward-modeling pipeline transforms intrinsic stream realizations into \LSST-like observed data. Indeed, it illustrates the impact of survey systematics on the spatial distribution of the point-like sources. It also highlights the impact on the stellar stream visibility, and the need for mitigation strategies aimed at improving the signal-to-noise ratio and reducing the impact of observational systematics.

\subsubsection{Injected vs. observed stellar stream}

Once the stream has been injected into the survey, a spatial cut is applied to define the `on-stream' region. This is particularly important because our analysis utilizes the one-dimensional linear density along the stream, making it sensitive to leakage from off-stream sources.
For a Gaussian signal in a uniform background, the aperture that maximizes the signal-to-noise ratio $S/\sqrt{B}$ corresponds to $|\phi_2| < 1.4\,\sigma$, where $\sigma$ is the dispersion of the Gaussian profile. Since the stream follows such a profile, we adopt this cut throughout the analysis. This choice has also been validated empirically at the level of gap detection efficiency.

The top panel of Fig.~\ref{fig:2Ddens} presents a stellar stream with a gap (see Sec.~\ref{sec:streammodeling}),  before (left) and after (right) injection at a given location in the \LSST footprint, after $4$ years of observation. The panels below show the linear density of the stream after applying the on-stream spatial selection described above.
The forward-modeling pipeline introduces inhomogeneities driven by depth variability, selection effects and background contamination, as illustrated by the darker patches in the upper-right panel of the two-dimensional density plot. These effects generate spurious density variations along the stream and reduce the contrast of intrinsic features. 
In particular, localized perturbations such as gaps can be significantly degraded or erased, as illustrated by the absence of strong evidence of the injected gap in the linear density (Fig.~\ref{fig:2Ddens}; top right panel).

\subsubsection{Mitigation strategy and additional cuts}
\label{subsubsec:birghtmagcut}

\paragraph{Mitigation of survey depth fluctuations}
Imposing a magnitude cut that is significantly brighter than \mlim is a simple strategy to mitigate the imprint of survey depth variations, but comes at the cost of a statistical loss.

The bottom panels of Fig.~\ref{fig:2Ddens} illustrate the impact that a bright cut on the magnitude has on the spatial distribution and stellar stream visibility.
In the left panel, restricting the sample to bright magnitudes restores a more homogeneous density along the stream (with respect to the top right panel).
This behavior arises because faint objects---especially compact galaxies---become increasingly numerous near the survey limit (see bottom panel in Fig.~\ref{fig:matchfilter}) and are strongly modulated by depth variations, imprinting artificial fluctuations in both the stream and background densities.
By removing these faint sources, the cut efficiently suppresses survey-induced inhomogeneities, while largely preserving the intrinsic density variations of the stream, such as gaps.

We therefore define a {\em position-dependent} cut designed to reduce survey-induced inhomogeneities. The threshold is defined as the 2nd percentile of the local $10\sigma$ magnitude limit distribution of the selected objects. This choice avoids relying on the most extreme local depth values, which could otherwise impose an unnecessarily strong cut if only a negligible fraction of the stream lies in unusually shallow regions. While not optimized, this prescription provides a conservative and realistic approximation of practical mitigation strategies. Typical distributions of these cuts are presented in Appendix~\ref{app:magcut}. The median cuts in the $g$ band are $24.2\magn$ for year 1 and $25.3\magn$ for year 4, while in the $r$ band they are $24.6\magn$ and $25.6\magn$, respectively. These values provide representative effective magnitude limits for realistic \LSST stream gap analyses accounting for survey depth mitigation. It has been found to be effective in reducing survey-induced systematics (see Sec. \ref{sec:results:sub:nobackground}), while leaving room for more refined optimization in future analyses.

\paragraph{Mitigation of S/G contamination}
Even in the absence of survey depth variations, a magnitude cut is necessary to control background contamination. In particular, imperfect star–galaxy classification leads to increasing galaxy leakage at faint magnitudes, which can dominate the selected sample.
To limit this, we define a magnitude threshold $m_g^{\mathrm{cut}}$ that maximizes the stream significance relative to the background,
\begin{equation}
    \mathrm{Sig}(m_g^{\mathrm{cut}}) = \frac{S(m_g^{\mathrm{cut}})}{\sqrt{B(m_g^{\mathrm{cut}})}} \, ,
    \label{eq:cut_snr}
\end{equation}
where $S$ and $B$ denote the number of selected stream stars and background objects, respectively. This selection reduces contamination from faint galaxies while retaining a sufficient number of stream stars for the analysis.

The bottom-right panel of Fig.~\ref{fig:2Ddens} shows an idealized configuration combining a uniform survey depth with perfect S/G, used as a benchmark for the realistic case in the bottom-left panel. Two systematics are absent here: depth is uniform, removing the need for the mitigation cut, and classification is perfect, removing galaxy leakage, dominant in the faint regime as shown in Fig.~\ref{fig:matchfilter} (bottom panel). A magnitude cut is still applied to maximize stream significance (Eq.~\ref{eq:cut_snr}), but is now driven solely by foreground-star contamination, allowing a fainter cut and higher contrast than in the realistic case (bottom-left panel). This illustrates that survey depth variations and imperfect classification each independently limit the usable survey depth.

In the presence of both survey depth variations and background contamination, we adopt the more restrictive of the S/N-based and depth-based magnitude cuts.

\subsection{Summary and discussion}
The \StreamObs framework, combined with the color-magnitude selection, provides a quick forward-modeling approach to assess the performance of photometric surveys for stellar stream studies. 
It consistently incorporates both statistical uncertainties and spatially varying systematics.

Mitigation strategies based on magnitude cuts are a necessary step to obtain robust stream characterization. Survey depth variations require masking parts of the data, following the same philosophy as in large-scale structure analyses, which also rely on relatively bright magnitude-limited samples to control observational systematics \citep{rodriguez-monroy:2025, weaverdyck:2026}. While more advanced correction schemes are being actively developed (e.g., \citealt{boone:2026}), they remain at the frontier of current analyses, and our approach therefore adopts a conservative choice.

Background contamination linked to star–galaxy classification imposes a fundamental limitation on the usable depth: galaxies begin to dominate beyond $m_g \simeq 24.5$, preventing a reliable exploitation of the full \LSST depth expected after 10 years ($m_g \sim 27$). While future improvements may recover part of this lost information, we adopt conservative cuts to ensure robustness in the present analysis.

\section{Mock observation setup}
\label{sec:protocol}

To quantify the impact of survey systematics on stream density fluctuation measurements, the analysis is anchored to an \AAU-like stream baseline, whose fixed parameters and their limitations are discussed in Sec.~\ref{sec:discussion}. Three classes of parameters are then varied independently: the intrinsic number of stream stars $N_\star^{\rm true}$ (Sec.~\ref{sec:injected_parameters}), to characterize the dependence of sensitivity on stellar statistics, which can also be interpreted as probing streams at different distances or surface brightnesses; the gap depth and width, to map the full detectability landscape over a realistic parameter range; and the survey systematics configuration (Sec.~\ref{subsec:popstudied}), to isolate and quantify the individual contribution of each observational effect to the overall sensitivity loss.

\subsection{Injected stream and gap parameters}
\label{sec:injected_parameters}

From the stellar stream gap model described in Sec.~\ref{sec:gap-model} and Eq.~\eqref{eq:gapmodel}, we generate $1500$ mock realizations for each gap parameter combination. We sample gap depths $A$ from $0$ (uniform stream) to $1$ (full depletion) in steps of $0.2$, and gap widths $w$ from $3\degree$ to $11\degree$ in steps of $2\degree$. The upper bound for $w$ is chosen to avoid unrealistically large gaps for streams with lengths comparable to \AAU. The gap position is randomly sampled, with the constraint that it lies fully within the stream realization as illustrated by Fig.~\ref{fig:mask}.

The number of true stars in the stream, before any observational effects, is varied from $2 \times 10^{4}$ to $6 \times 10^{4}$ in steps of $10^{4}$. 
This range is chosen to bracket realistic stream populations accessible to \LSST observations (see discussion in Sec.~\ref{sec:Nstar_norm}). We summarize the correspondence between the injected number of stream stars, the associated surface brightness, and stellar mass in Table~\ref{tab:stream_properties}. We also estimate the number of stars recovered in uniform \LSST observations after 1 and 4 years, corresponding respectively to $\sim10.3\%$ and $\sim14.4\%$ of the injected population. These values are derived for a homogeneous survey without dust extinction, and therefore vary with stream position in realistic survey configurations due to depth fluctuations and dust absorption.

\begin{table}
\centering
\caption{Reference properties of the sampled \AAU-like stream realizations used throughout this work. For each injected number of stars $N^{\rm true}_\star$, we list the corresponding stellar mass, observed surface brightness, and the average number of observed stars recovered in uniform \LSST Y1 and Y4 configurations. The bold row corresponds to the baseline \AAU-like configuration adopted throughout the analysis.}

\label{tab:stream_properties}
\begin{tabular}{ccccc}
\hline
$N^{\rm true}_\star$ &
$M_\star^{\rm true}$ &
$\Sigma_\star$ &
$N^{\rm obs}_\star$ &
$N^{\rm obs}_\star$ \\
&
$[M_\odot]$ &
$[\mathrm{mag/arcsec^2}]$ &
(Y1) &
(Y4) \\
\hline
$2\times10^4$ & $4.87\times10^3$ & $33.66$ & $2079$ & $2868$ \\
$3\times10^4$ & $7.31\times10^3$ & $33.22$ & $3086$ & $4312$ \\
$\mathbf{4\times10^4}$ & $\mathbf{9.74\times10^3}$ & $\mathbf{32.90}$ & $\mathbf{4139}$ & $\mathbf{5750}$ \\
$5\times10^4$ & $1.22\times10^4$ & $32.66$ & $5143$ & $7169$ \\
$6\times10^4$ & $1.46\times10^4$ & $32.46$ & $6196$ & $8628$ \\
\hline
\end{tabular}
\end{table}

\subsection{Survey systematics configurations}
\label{subsec:popstudied}

The impact of survey systematics is assessed on the configurations listed in Table~\ref{tab:configurations}. We explore configurations ranging from idealized to realistic (some of which are highlighted in Fig.~\ref{fig:2Ddens}).
The survey with no depth variation and no background is chosen as the idealized reference scenario, with the gap recovery limited only by the Poisson statistics of the detected number of stars (ideal case with infinite S/N ratio). The star counts are limited by the object detection efficiency (Eq.~\ref{eq:eff_detect}), and not by stellar classification (Eq.~\ref{eq:eff_class}), which is considered to be perfect in this ideal case. It closely matches the approach of some dynamical simulation studies, where the observed stream population is obtained by applying a simple magnitude cut around the survey limiting magnitude \citep{erkal:2015a,lu:2025,nguyen:2025}.

A slightly more realistic configuration is to add realistic star classification efficiency (Sec.~\ref{sec:star_detect_efficiencies}), 
background contamination (Sec.~\ref{sec:bkdg}) from foreground stars and misclassified galaxies (Sec.~\ref{sec:galaxy_missclassification}), as well as survey depth inhomogeneities (Sec.~\ref{sec:depth_maps}). When including the background, a signal-to-noise-based magnitude cut is applied to control the galaxy contamination. For inhomogeneous surveys, an additional depth-based cut (Sec.~\ref{subsubsec:birghtmagcut}) mitigates survey-depth fluctuations; both cases are tested to isolate the background impact.
The configuration that we consider the most realistic (i.e., representative of the expected data from \LSST) combines all of these effects.

\begin{table}
\centering
\caption{Configurations of simulated datasets and associated selection criteria, ordered from ideal (in {\em italics}) to the most realistic (in {\bf boldface}).}
\label{tab:configurations}
\begin{tabular}{lcccc}
\hline
Survey  &\!\!\!\!\!\!Background\!\!\!\!\!\!& Magnitude &\!\!\!\!\!\! S/G \!\!\!\!\!\\
depth   &\!\!\!\!\!\!inclusion\!\!\!\!\!\!& selection  &\!\!\!\!\!\!classification\!\!\!\!\!\\
\hline
\multirow{2}{*}{{\em Homogeneous}}  & \multirow{2}{*}{{\em No}}  &  \multirow{2}{*}{{\em None}} & \hspace{-0.5cm}\rdelim\{{2}{0.4cm}{\em Perfect}\\
   &&& Observed\\[2mm]
\multirow{2}{*}{Realistic}    & \multirow{2}{*}{No}  & \multirow{2}{*}{None} & \hspace{-0.5cm}\rdelim\{{2}{0.4cm}Perfect \\
   &&& Observed\\[2mm]
\multirow{2}{*}{Realistic}    & \multirow{2}{*}{No}  & \multirow{2}{*}{Depth-based cut} & \hspace{-0.5cm}\rdelim\{{2}{0.4cm}Perfect\\
   &&& Observed\\[2mm]
\multirow{2}{*}{Homogeneous}  & \multirow{2}{*}{Yes} &  Matched filter \& &  \hspace{-0.5cm}\rdelim\{{2}{0.4cm}Perfect\\
        & & S/N-based cut & Observed\\[2mm]
\multirow{2}{*}{Realistic}  & \multirow{2}{*}{Yes} &  Matched filter \& &  \hspace{-0.5cm}\rdelim\{{2}{0.4cm}Perfect\\
        & & S/N-based cut & Observed\\[2mm]
\multirow{2}{*}{\textbf{Realistic}}    & \multirow{2}{*}{\textbf{Yes}} & \textbf{Matched filter \&} & \hspace{-0.5cm}\rdelim\{{3}{0.4cm}Perfect\\
&&  {\bf S/N-based cut \&}&\\
&& {\bf Depth-based cut}& {\bf Observed}\\[2mm]
\hline
\end{tabular}
\end{table}

\section{Statistical framework for gap detection}
\label{sec:analysis}
In order to quantify the effect of survey systematics on the measurement of density fluctuations in stellar streams, we need to specify a sound metric to assess the detection efficiency of the injected gaps.

\subsection{Power spectrum analysis and likelihood}

Several DM studies with stellar streams in the literature have been performed in Fourier space  \citep{bovy:2017, banik:2021}. This representation provides a convenient framework to quantify density fluctuations across a wide range of spatial scales. It has been shown to be well suited to characterizing the impact of observational systematics \citep{boone:2026}.
An additional advantage of the Fourier approach is that it naturally captures fluctuations over all scales simultaneously. In contrast to real-space methods, which are typically sensitive to localized and visually prominent gaps, the power spectrum encodes the cumulative effect of both small- and large-scale perturbations. This makes it especially sensitive to the global fluctuation level induced by survey effects, rather than focusing on individual features in isolation.
In this study, we use both approaches, i.e. analysis done in real and Fourier space, in order to quantify the impact of this choice of metric on the results. They yield consistent results (App.~\ref{app:poisson}), with the Fourier analysis being slightly more conservative, and we therefore adopt it throughout the rest of the paper.

\paragraph{Likelihood in Fourier space}

The one-dimensional linear density $\lambda_n$ is binned, see Eq.~\eqref{eq:lindens_model}, in intervals of $0.5\degree$,  within a window extending $2\degree$ beyond each end of the stream, fully encompassing it. The power spectrum of the data  $P_d(k)$ and the model  $P_m(k)$ (Sec.~\ref{sec:gap_model}) are computed using \code{scipy.signal.csd}. When relevant (see next section), a constant background contribution is added in the model to account for a uniform contamination level.

The statistical properties of the power spectrum are modeled using a likelihood based on the non-central $\chi^2$ distribution,  $\chi^2_{\rm nc}$. 
For each Fourier mode $k > 0$, the likelihood is written as
\begin{equation}
    \mathcal{L} = \prod_{k>0} 
    \chi^2_{\mathrm{nc}}\left(P_d(k)\,;\, \mathrm{df}=2,\ \lambda,\sigma(k) \right) \, ,
    \label{eq:likelihood}
\end{equation}
where $\mathrm{df}$ is the number of degrees of freedom, $\lambda$ is the non-centrality parameter, and $\sigma(k)$ is a scale parameter that encodes the Poisson noise from the available objects sample.

The $\chi^2_{\rm nc}$ distribution generalizes the standard $\chi^2$ distribution by introducing a non-centrality parameter, $\lambda$, which encodes the presence of a signal \citep{Patnaik:1949}. 
In the limit that $\lambda = 0$, it reduces to a central chi-squared distribution.

In the absence of signal, we recover that the power spectrum $P_d(k)$ of a Gaussian random field is a central $\chi^2$ with  $\mathrm{df} = 2$ \citep[Sec.~6.1.3]{priestley1981spectral}; see also \citet[Sec.~5]{vanderplas:2018}. This allows us to estimate the scale parameter $\sigma(k)$ by fitting the distribution of $P_d(k)$ in noise-only realizations.
In the presence of a signal, the power spectrum distribution becomes non-central.
Empirically, the non-centrality parameter is found to scale as
\begin{equation}
    \lambda(k) = \frac{P(k)}{\sigma(k)} \, ,
\end{equation}
where $P(k)$ denotes the signal contribution to the power spectrum. 
The $\chi^2_{\rm nc}$ parametrization thus provides an effective description of the transition from noise-dominated to signal-dominated regimes. 

In the context of a likelihood-ratio test for nested models (see Sec.~\ref{sec:hypothesis}), the likelihood-ratio distribution is asymptotically described by a central $\chi^2$ distribution under the null hypothesis $H_0$ (no signal) and a non-central $\chi^2$ distribution under the alternative hypothesis $H_1$ \citep[see Sec.~10.5.2 of][]{2006smep.book.....J}.\footnote{In our context, $H_0$ and $H_1$ correspond to the unperturbed and perturbed stream, respectively.} Moreover, the non-centrality parameter, $\lambda$, is given by the squared signal significance in units of standard deviations (analogous to signal-to-noise ratio). We verified both of these behaviors using idealized simulations, giving us confidence in our choice of the likelihood function Eq.~\eqref{eq:likelihood}.

This likelihood neglects, however, correlations between Fourier modes induced by the finite stream length. We nevertheless tested this approximation in a controlled case using sinusoidal density variations, instead of the single gap model discussed in Sec.~\ref{sec:gap-model}, which produces localized Fourier features, and found consistent likelihood fitting results (see also App.~\ref{app:poisson} for a comparison with the gaps model).

\subsection{Hypothesis testing}
\label{sec:hypothesis}

The likelihood framework is used to perform hypothesis testing on the presence of density fluctuations in stellar streams. 
The null hypothesis $H_0$ corresponds to a uniform stream, characterized by a flat density profile and thus an amplitude of fluctuations $A = 0$.
The alternative hypothesis $H_1$ describes a perturbed stream with $A > 0$, including additional parameters that depend on the chosen model, for instance the width $w$ and position $x_{\mathrm{gap}}$ of the gap.

The compatibility between the data and these hypotheses is quantified using the delta-log-likelihood \citep{cowan:2011},
\begin{equation}
    R = -2 \left[ \log \mathcal{L}_{H_0}(A=0, \theta) - \log \mathcal{L}_{H_1}(A, \theta) \right] \, ,
    \label{eq:likelihoodratio}
\end{equation}
where $\theta$ denotes the set of additional parameters associated with the model. This statistic measures the relative preference for a fluctuating density model over a uniform stream.

\begin{figure}[t]
\centering
\includegraphics[width=\columnwidth]{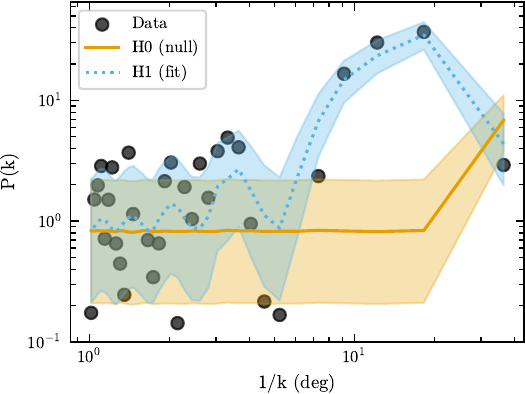}
\caption{Example of a one-dimensional power spectrum fit for a stellar stream with $N^{\mathrm{true}}_\star  = 2 \times 10^{4}$ stars, observed after 1 year of \LSST operation. 
The realization (symbols) includes a gap with amplitude $A = 0.5$ and no background contamination.
Shaded regions indicate the $16$th and $84$th percentiles of the distribution arising from Poisson uncertainties, while dashed and solid lines show the corresponding medians.}
\label{fig:pk}
\end{figure}

Figure~\ref{fig:pk} shows an example of the power spectrum fit and the corresponding hypothesis comparison.
Under $H_0$, the power spectrum exhibits an increase at large scales due to the finite stream length, and approaches a constant level at small scales where shot noise dominates. 
Under $H_1$, the model reproduces the excess power induced by the gap at large scales. Residual features at smaller scales arise from windowing effects in the Fourier transform. Comparing both hypotheses results for this realization yields $R\sim170$.
At small scales, we observe a noise-dominated distribution that follows a $\chi^2$ behavior, which is intrinsically asymmetric. As a result, percentile-based intervals are not symmetric around the median, and some realizations may appear to lie outside the displayed bands due to visualization effects rather than model inadequacy.

\subsection{Detection efficiency}
The detection efficiency of density fluctuations is quantified from the distribution of the likelihood ratio defined in Eq.~\eqref{eq:likelihoodratio}.

For a given set of gap parameters (depth and width; Eq.~\ref{eq:gapmodel}) and survey configuration (Table~\ref{tab:configurations}), $1500$ stream realizations are generated and injected within the \LSST footprint (see Sec.~\ref{sec:injected_parameters}), providing the sample on which the hypothesis test described below is performed.

\paragraph{Null hypothesis distribution.}

The reference distribution is obtained under the null hypothesis $H_0$, corresponding to a uniform stream.
Its likelihood ratio distribution, $R(A=0)$, defines the baseline for detection.
The $95$th percentile of this distribution is adopted as the rejection threshold, corresponding to a false positive rate of detecting stream density fluctuations of $5\%$.

The null distribution depends on the observational configuration, including background contamination and selection cuts (see Table~\ref{tab:configurations}). 
It therefore captures the impact of survey systematics, which introduce artificial density fluctuations along the stream. These effects shift the likelihood ratio toward larger values and broaden its distribution, requiring an increased rejection threshold in order to achieve the same 5\% false positive rate.
Consequently, the efficiency is reduced, since genuine fluctuations must exceed the level of spurious structure induced by the survey in order to be detected. 

\begin{figure}[t]
\centering
\includegraphics[width=\columnwidth]{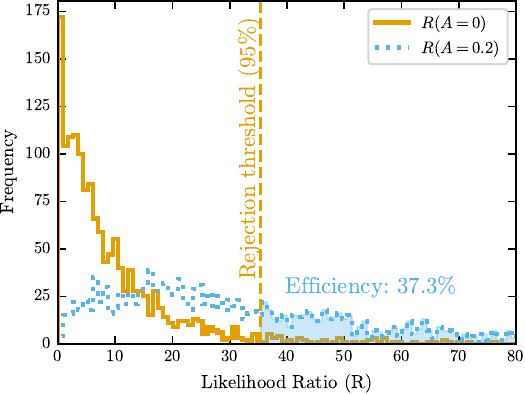} \caption{Distributions of the likelihood ratio for realizations of a stellar stream with and without a gap. This example corresponds to $N^{\mathrm{true}}_\star  = 2 \times 10^{4}$ stars, without background contamination, after 4 years of \LSST observation.
}
\label{fig:R0vsR}
\end{figure}

Figure~\ref{fig:R0vsR} shows an example of the likelihood ratio distributions for the null hypothesis in orange full line, and the resulting rejection threshold. 
The $H_0$ distribution exhibits the expected chi-squared-like behavior. 
As an additional validation of the likelihood pipeline, it is found to be well described by a $\chi^2$ component and Dirac at 0, consistent with Chernoff's  extension of Wilks' theorem to bounded parameters \citep{chernoff:1954, wilks:1938}.

\paragraph{Signal hypothesis and efficiency.}
For a given fluctuation model with parameters $\theta$ (e.g., gap depth $A$ and width $w$), the same procedure is applied to generate the likelihood ratio distribution $R(A,\theta)$.
A stream realization is considered to be detected if its likelihood ratio exceeds the rejection threshold defined from the null hypothesis distribution.

The detection efficiency is then defined as
\begin{equation}
{\cal E}(\theta) =
\frac{n\!\left[R(A,\theta) > R^{95\mathrm{th}}(A=0)\right]}
{N_{\rm real}},
\label{eq:eff}
\end{equation}
where $n[\cdot]$ counts realizations satisfying the condition, and $N_{\rm real}=1500$ is the total number of realizations per stream configuration.

When a signal is present, the distribution is shifted toward larger values of $R$, reflecting the reduced compatibility of the data with the uniform model ($H_0$). 
In the example shown in Fig.~\ref{fig:R0vsR}, corresponding to a gap with amplitude $A = 0.2$, $37.3\%$ of the realizations exceed the rejection threshold. This defines the detection efficiency for this stream type and survey configuration.

\section{Gap detectability forecast in \LSST survey}
\label{sec:results}

With the forward-modeling pipeline (Sec.~\ref{sec:genobs}), the stream model (Sec.~\ref{sec:streammodeling}), and the statistical framework (Sec.~\ref{sec:analysis}) in place, the detectability of gaps (Sec.~\ref{sec:results:sub:nobackground} and \ref{sec:results:sub:wbackground}) is now assessed for each survey configuration (Sec.~\ref{subsec:popstudied}). We then comment on the impact of systematics on the minimal DM subhalo mass probed (Sec.~\ref{sec:subhalomass}).

\subsection{Ideal case: survey with no background}
\label{sec:results:sub:nobackground}

The background-free configuration provides a controlled setup to validate the pipeline and to interpret the impact of survey systematics in the absence of contamination. By construction, this configuration does not need spatial cuts for the stream selection.

\begin{figure*}[t]
\centering
\includegraphics[width=\textwidth]{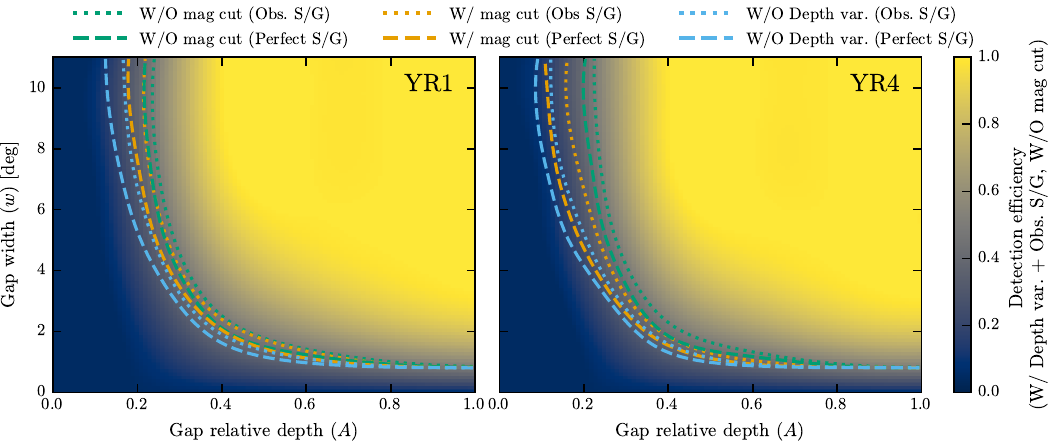}
\vspace{0.5cm}
\includegraphics[width=\textwidth]{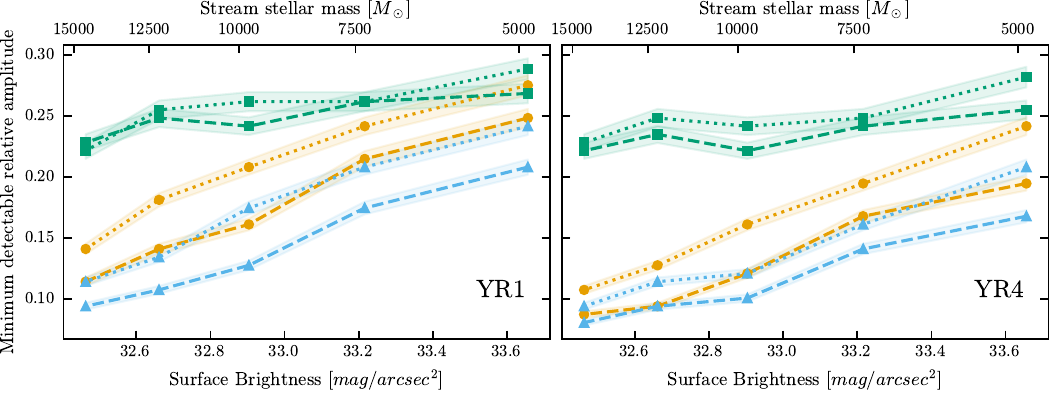}
\caption{
Detection performance for gap signatures in the absence of background contamination.
The upper panels show the detection efficiency as a function of gap depth $A$ and width $w$ for  $N^{\mathrm{true}}_\star=4 \times 10^4$.
The color scale indicates the efficiency, assuming a star-galaxy classification based on the extendedness parameter and no bright magnitude cut.
Contours mark the ${\cal E} (A,w) = 0.5$ level for different analysis methods.
The lower panels present the minimum detectable gap depth at fixed width $w=5\deg$, defined by a detection efficiency ${\cal E} > 0.5$, as a function of stream stellar mass (top scale) and surface brightness (bottom scale).
From left to right, the markers correspond to $N^{\rm true}_\star= (6,\,5,\,4,\,3,\,2)\times 10^4$ stars.
 Results are shown for 1 year (left) and 4 years (right) of observation.}
\label{fig:efficiency_nobackground}
\end{figure*}

Figure~\ref{fig:efficiency_nobackground} illustrates this impact for \LSST 1 year (left panels) and 4 years (right panels) on (i) the detection efficiency ${\cal E}$ as a function of the gap depth $A$ and width $w$ (top panels) and, in a complementary view, on (ii) the minimal detectable gap depth as a function of the number of stars in the stream as represented by the surface brightness and stellar mass (bottom panels).

\paragraph{Configuration sensitivity vs.\ $A$ and $w$}
In the top panels the detection efficiency ${\cal E}(A,\,w,\,N^{\rm true}_\star=4\times\,10^4)$, color-coded from 0 to 1 (from blue to yellow), increases with both $A$ and $w$, as deeper and wider gaps produce stronger and more extended signatures in the density field. The same trend is observed for the other background-free configurations.

The impact of the survey configurations is illustrated by drawing, on the same figure, the iso-contours ${\cal E}(A,\,w) = 0.5$ for each configuration. Because the efficiency is estimated from a finite number of realizations (see Sec.~\ref{sec:analysis}), bootstrap resampling gives a sampling uncertainty of $\sim6\%$ at ${\cal E}=0.5$, corresponding to an absolute uncertainty of $\sim0.03$.

Moving from a perfect (dashed lines) to observed (dotted lines) star-galaxy classification always degrades the detectable amplitude and width (the iso-curves move towards the upper right direction). The {\em ideal} configuration (blue lines) shows the smallest gaps that can be detected over the Poisson fluctuations. This gap sensitivity is degraded (iso-curves shifted to the top right) in inhomogeneous surveys (green lines), but enforcing a magnitude cut allows recovering part of this sensitivity (yellow lines).

At $w=0$, the gap modulation in Eq.~\eqref{eq:gapmodel} reduces exactly to a uniform stream, independently of $A$; below the $0.5\degree$ binning scale (Sec.~\ref{sec:analysis}), the gap is also unresolved. Both effects drive every configuration toward the same low efficiency at small $w$, explaining the convergence of the iso-curves there.

\paragraph{Configuration Sensitivity for 1 year vs.\ 4 years }
The most striking feature of Fig.~\ref{fig:efficiency_nobackground} is that, without a magnitude cut (the color scale shown), the efficiency maps (top panels) are nearly identical between the 1-year and 4-year surveys.
The larger number of detected stars in the deeper survey is therefore offset by an increased sampling of residual depth fluctuations, maintaining a comparable level of artificial structure between the two releases. This effect is quantified below at fixed gap width: without a magnitude cut, the minimum detectable gap depth improves by only $\sim3$--$6\%$ on average between the two releases, whereas once mitigation cuts are applied, the improvement instead reaches $\sim25$--$35\%$, comparable to the gain observed in uniform survey configurations. This shows that controlling survey-induced fluctuations is necessary to effectively benefit from the increased statistical power of deeper \LSST observations.

\paragraph{Configuration sensitivity vs.\ $A$ and $N^{\rm true}_\star$}
The bottom panels of Fig.~\ref{fig:efficiency_nobackground} show the minimum detectable relative gap depth as a function of the stream surface brightness. This quantity, denoted $A_{\mathrm{min}}$, is defined as the smallest gap depth for which the detection efficiency reaches $50\%$ at a fixed gap width of $w = 5^{\degree}$,
\begin{equation}
    {\cal E}(A_{\mathrm{min}},\,N^{\rm true}_\star,\,w=5^{\degree}) = 0.5 \, .
\end{equation}
The finite sampling uncertainty on the efficiency propagates into an uncertainty of $\sim3\%$ on $A_{\mathrm{min}}$, which is adopted throughout the rest of this work.

The ranking (between configurations) of the minimal detectable amplitude remains the same for the various stellar stream masses.
Unsurprisingly, the gap sensitivity decreases (i.e., the minimal detectable gap amplitude increases) when the surface brightness increases---or equivalently, when the stream mass (or $N_\star^{\rm true}$) decreases. The only exception is for the inhomogeneous survey without magnitude cuts (green squares), for which the sensitivity shows only a weak improvement when increasing the stellar mass. In this case, the spatial variations of the survey depth imprint artificial density fluctuations along the stream (false positives). As the stellar mass increases, the stream more densely samples these variations, enhancing the correlation between the observed density and the depth pattern.
Statistically speaking, the likelihood ratio, Eq.~\eqref{eq:likelihoodratio}, increases not only for perturbed streams but also for unperturbed stream realizations.
The null hypothesis distribution is therefore broadened and shifted toward larger values, raising the rejection threshold.
This effect compensates the gain in statistical power expected from larger samples, and limits the improvement in sensitivity. This balance is also reflected by the close agreement between the two green curves, corresponding to observed and perfect star-galaxy classification. Improving the star-galaxy classification alone does not significantly enhance the sensitivity, indicating that the regime is already dominated by survey depth variations rather than statistical noise.

This issue can be mitigated by applying magnitude cuts (orange circles), allowing us to recover a minimum detectable gap $A\sim 10\%$ deeper.

\paragraph{In brief}
In an (unrealistic) background-free survey, mitigating the impact of depth inhomogeneity is essential.
A simple and effective approach is to apply a bright magnitude cut adapted to the local depth, as defined in Sec.~\ref{subsubsec:birghtmagcut}.
In this case, the minimum detectable gap depth becomes limited by Poisson statistics rather than by survey-induced fluctuations.
This also explains the improved performance obtained with perfect star-galaxy classification.
This approach also restores the benefit of deeper observations (4 years).

\subsection{Realistic case: survey with background}
\label{sec:results:sub:wbackground}
We repeat the analysis for the survey configurations with background listed in Table~\ref{tab:configurations}. 

\begin{figure*}[t]
\centering
\includegraphics[width=\textwidth]{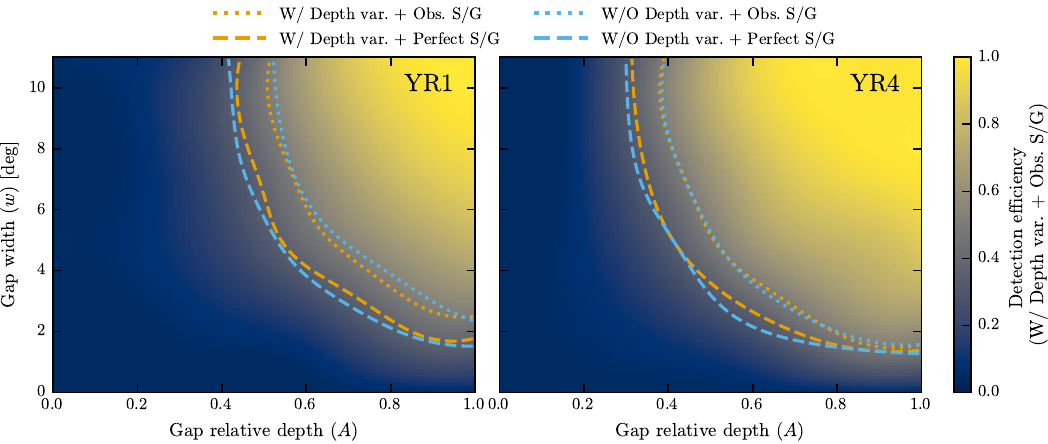}
\caption{Same as the top panel of Fig.~\ref{fig:efficiency_nobackground}. It shows the gap sensitivity detection in various scenarios, in the presence of background contamination. The color map refers to the results for the most realistic survey configuration only (inhomogeneous survey, with background and S/G contamination).}
\label{fig:efficiency_wbackground}
\end{figure*}

\paragraph{Comparison of performances w/ and w/o background}

A comparison of the top panels of Figs.~\ref{fig:efficiency_nobackground} (no background) and~\ref{fig:efficiency_wbackground} (with background) shows that, in both cases, the detection efficiency ${\cal E}(A,\,w,\,N^{\rm true}_\star=4\times\,10^4)$ increases from ${\cal E}=0.05$ (blue) to ${\cal E}=1$ (yellow) with deeper and wider gaps $(A,w)$; the iso-contours for different survey configurations (w/ or w/o depth variation and star classification) also display similar dependences.
However, with background, the sensitivity is significantly reduced: iso-contours ${\cal E}(A,\,w) = 0.5$ are shifted significantly towards deeper and wider gaps. 
Another significant difference is between the perfect and observed star-galaxy classification configurations (compare the dashed and dotted lines in the two figures): in addition to the star-galaxy classification, present in the background-free configuration, the background configuration adds numerous misclassified galaxies in the stellar stream selection.
A last difference is the increased improvement for the 4-year survey over the 1-year survey in the background case: this arises from the effective mitigation of systematics or signal-to-noise ratio magnitude cuts (discussed in Sec.~\ref{subsubsec:birghtmagcut}).

\subsection{Overall sensitivity loss in realistic surveys}
\label{sec:ideal_vs_realistic}

\begin{figure}[t]
\centering
\includegraphics[width=\columnwidth]{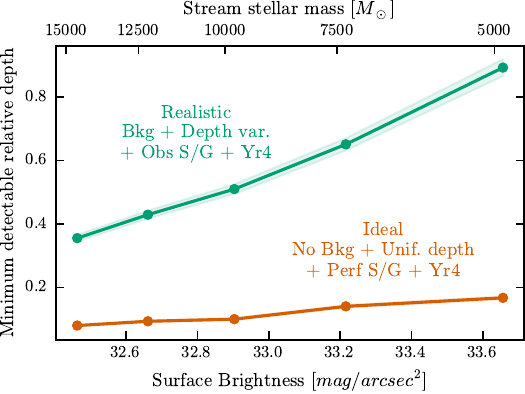}\\
\includegraphics[width=\columnwidth]{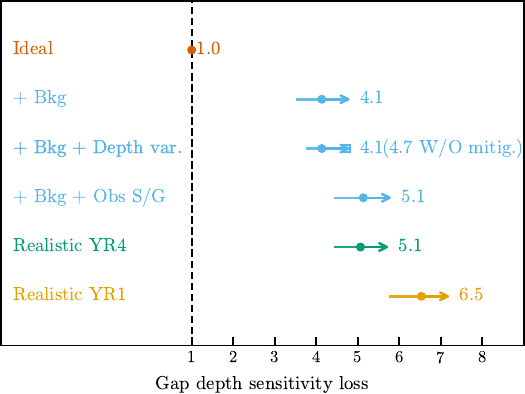}
\caption{The top panel is similar to that of Fig.~\ref{fig:efficiency_nobackground}. It shows the minimal detectable relative amplitude, obtained from the efficiency condition ${\cal E}(A^{\rm min},w=5\deg)=0.5$, as a function of stream stellar mass (top axis) and surface brightness (bottom axis) after 4 years of observation. For readability, we only show the result for the realistic (green lines) and ideal (orange lines) survey configurations.
The bottom panel presents the ratio $A^{\mathrm{min}} / A^{\mathrm{min}}_{\mathrm{ideal}}$, highlighting the degradation induced by different survey systematics. Markers indicate the reference case with $N^{\mathrm{true}}_\star = 4 \times 10^{4}$ stars, while error bars and arrows reflect the dependence on surface brightness. The square marker indicates the case without survey depth variation mitigation.}
\label{fig:gap_sensitivity}
\end{figure}

We show in the top panel of Fig.~\ref{fig:gap_sensitivity}, the  minimal detectable gap relative depth---from  ${\cal E}(A_{\mathrm{min}},\,N^{\rm true}_\star,\,w\!=\!5^{\degree})=0.5$---as a function of the stream surface brightness. We only show the two extreme (ideal and realistic) configurations of Table~\ref{tab:configurations}.
The realistic survey (green lines) exhibits a much stronger dependence with the surface brightness, compared to the ideal survey (orange lines). This pronounced dependence on the number of stars, which was not present in the background-free case, reflects the dominant role of background contamination.

The bottom panel summarizes our results, by showing the ratio of the minimum detectable depth (for various survey systematics) to the {\em ideal} survey configuration $A_{\mathrm{min}}/A_{\mathrm{min}}^{\mathrm{Ideal}}$. For the 4-year survey,
\begin{itemize}
    \item the gap depth detectability degrades by a factor $\sim 4$ when adding the background, even for a homogeneous depth and perfect S/G classification;
    \item applying bright magnitude cuts to mitigate survey depth variations recovers $\sim15\%$ of the sensitivity;
    \item depth spatial variation and S/G classification further degrades the sensitivity by $\sim 23\%$ (i.e., factor $\sim 5$ sensitivity loss with respect to the ideal survey);
    \item for a 1-year survey, the sensitivity loss for the realistic survey is $\sim 30\%$.
\end{itemize}

\subsection{Minimal detectable subhalo mass}
\label{sec:subhalomass}

We can connect our stream gap sensitivity studies to DM subhalo sensitivity by mapping the depth and width of a gap to the mass of a DM subhalo that could produce it.  

\paragraph{Probability to produce and detect a gap}

Following the analytical framework of \cite{erkal:2016}, the conditional probability $P(A,w \mid M_{\mathrm{sub}})$, of the gap property distribution $(A,w)$ given a subhalo mass $M_{\mathrm{sub}}$, is obtained by marginalizing over parameters like the impact time and velocity (see App.~\ref{app:subhalosproba}).
This mapping, while approximate (compared to full dynamical simulations), is sufficient to assess how observational systematics affect the minimal detectable subhalo mass. 

This conditional probability enables us to write the probability that a subhalo with mass $M_{\mathrm{sub}}$ produces a detectable gap (provided an interaction with the stellar stream occurred):
\begin{equation}
    P^{\rm detect}(M_{\mathrm{sub}}) =
    \int \int {\cal E}(A,w)\,
    P(A,w \mid M_{\mathrm{sub}})\,
    \mathrm{d}A\,\mathrm{d}w \, ,
    \label{eq:proba_subhalM}
\end{equation}
where ${\cal E}(A,\,w,\, N^{\rm true}_\star=4 \times 10^4)$ is the gap detection efficiency estimated in the previous section.
We stress that there is no assumption on the abundance or spatial distribution of subhalos in this calculation, therefore it should not be interpreted as a detection rate.

\begin{figure}[t]
\centering
\includegraphics[width=\columnwidth]{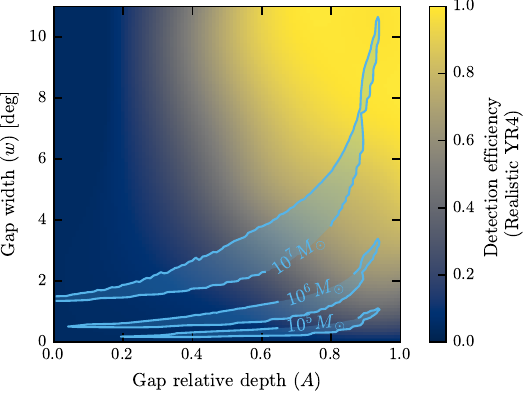}
\caption{68\% highest-density contours of the gap distributions induced by subhalos of $10^5$, $10^6$ and $10^7\,M_\odot$ (from bottom to top) in the $(A,w)$ plane. The background color map indicates the gap detection efficiency for a realistic four-year \LSST survey observing a stellar stream with $N^{\mathrm{true}}_\star = 4 \times 10^{4}$ stars.}
\label{fig:subhalo_distribcontours}
\end{figure}

\begin{figure}[t]
\centering
\includegraphics[width=\columnwidth]{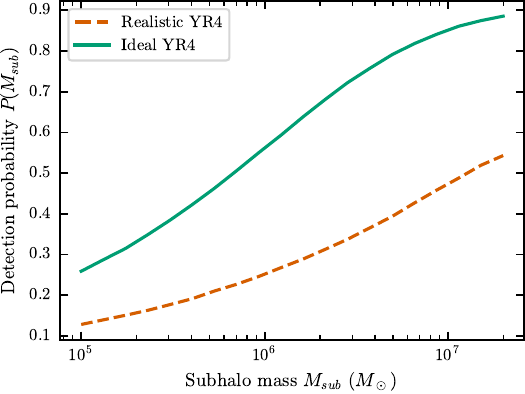}
\caption{Detection probability of DM subhalos as a function of their mass, for a stellar stream with $N^{\mathrm{true}}_\star= 4 \times 10^{4}$ stars observed with \LSST after 4 years. The ideal case corresponds to streams observed without background and survey systematics effects, while the realistic one includes those effects.}
\label{fig:subhalo_proba}
\end{figure}

The two terms of this integral are shown in Fig.~\ref{fig:subhalo_distribcontours}. The displayed contours correspond to the minimum-area regions enclosing 68\% of the probability density. More massive subhalos tend to produce larger gaps, their distribution encompassing regions of higher detection efficiency, therefore increasing their detectability.
We show the resulting detection probability $P^{\rm detect}$ in Fig.~\ref{fig:subhalo_proba}, for both ideal and realistic survey configurations (see Table~\ref{tab:configurations}). As expected, the probability increases with subhalo mass. Overall, the realistic survey configuration yields up to $\sim 2$ times fewer detectable gaps than the ideal case.
Additional configurations (not shown) indicate that depth fluctuations alone reduce the detection probability by $\sim 9\%$, while imperfect star-galaxy classification leads to a larger $\sim 24\%$ decrease.
The realistic 1-year survey further decreases the detection probability by a factor of $\sim 0.70$ relative to the 4-year configuration.

\paragraph{Minimal subhalo mass sensitivity}

To summarize the sensitivity of the survey, we define a characteristic mass scale by requiring $P^{\rm detect}(M_{\mathrm{sub}}) > 0.4$.\footnote{The conservative threshold of 0.4 is chosen for practical reasons: some realistic configurations, especially the 1-year \LSST survey, do not reach probabilities much larger than $P^{\rm detect}(M_{\mathrm{sub}})\sim0.4$ for the subhalo mass range of interest; adopting $P^{\rm detect}=0.5$ leads to stronger inferred sensitivity losses from background contamination and survey systematics and reduces the minimum detectable subhalo mass by a factor of $\sim2$.}
This threshold identifies the mass above which a non-negligible fraction of subhalo impacts produce detectable signatures.
It should be interpreted as an intrinsic detectability limit of the measurement, rather than a prediction for the number of observed subhalos, since no assumption is made on the underlying subhalo mass function.

\begin{figure*}[t]
\centering
\includegraphics[width=0.52\textwidth]{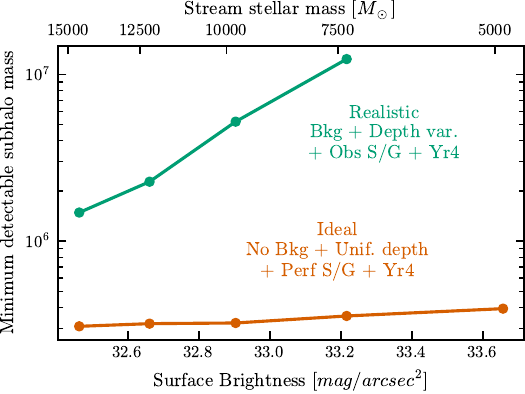}
\includegraphics[width=0.46\textwidth]{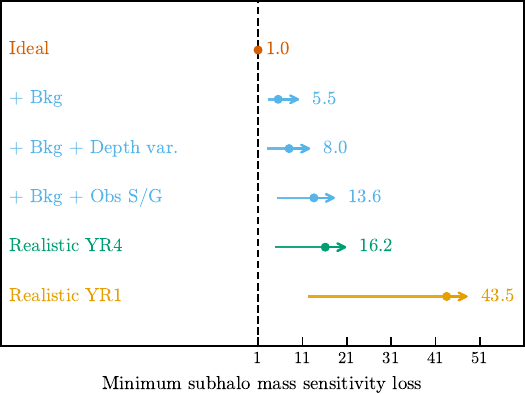}
\caption{(Left) The minimum detectable subhalo mass defined by $P(M_{\mathrm{sub}}) > 0.4$, as a function of stream stellar mass (top scale) and surface brightness (bottom scale) for 4 years of observations, comparing realistic and ideal survey configurations.
(Right) The ratio of the minimum detectable subhalo mass relative to the ideal survey $M^\mathrm{min}/M^\mathrm{min}_\mathrm{ideal}$, for different systematics consideration.
Markers indicate the reference case with $N^{\mathrm{true}}_\star  = 4 \times 10^{4}$ stars, while error bars and arrows reflect the variation with the number of stars in the stream.}
\label{fig:subhalosensi}
\end{figure*}

Figure~\ref{fig:subhalosensi} (left panel) shows the minimum detectable subhalo mass as a function of the stream surface brightness or stellar mass. Configurations without displayed points correspond to cases where the detection probability does not reach the adopted threshold within the explored mass range, typically for the faintest streams.

As for gap detectability, the realistic survey configuration exhibits a much stronger dependence on surface brightness than the idealized case. In the ideal configuration, the minimum detectable mass reaches values nearly one order of magnitude smaller than in the realistic survey, and depends only weakly on the number of stars. This behavior reflects the transition from a Poisson-noise-dominated regime in the ideal case to a background-dominated regime in the realistic configuration, where contamination and survey systematics significantly degrade the signal-to-noise ratio.

The right panel of Fig.~\ref{fig:subhalosensi} summarizes the relative degradation in subhalo mass sensitivity with respect to the ideal survey configuration. For a representative stream with $N^{\mathrm{true}}_\star = 4 \times 10^{4}$ stars,
\begin{itemize}
    \item background contamination alone increases the minimum detectable subhalo mass by a factor of $\sim 5.5$;
    \item survey depth variations further degrade the sensitivity to a factor of $\sim 8$, mainly due to the mitigation cuts required to suppress spurious fluctuations;
    \item imperfect star-galaxy classification increases the minimum detectable mass by a factor of $\sim 13.6$ relative to the ideal case;
    \item a realistic 1-year survey further reduces the sensitivity by a factor of $\sim 2.7$ compared to the 4-year configuration.
\end{itemize}
Hence, while background contamination is the dominant limitation, survey systematics further degrade the subhalo mass sensitivity by an additional factor of $\sim 2.9$ relative to the background-only configuration, driven primarily by imperfect star-galaxy classification.

Overall, these results demonstrate that observational effects---background contamination, survey inhomogeneities, and classification uncertainties---can shift the minimum detectable subhalo mass by more than an order of magnitude. Since the abundance of DM subhalos rises steeply toward low masses ($dN/dM \propto M^{-1.9}$ \citep{springel:2008}), this loss of sensitivity directly translates into a reduced ability to test deviations from $\Lambda$CDM. Accurately modeling these effects is therefore essential for any realistic forecast of DM constraints from stellar streams.

\section{Discussion}
\label{sec:discussion}

\subsection{Analysis choices}
Several of our analysis choices were adopted to provide a consistent and controlled analysis framework rather than an optimized analysis strategy.
Overall, these choices do not affect the qualitative trends highlighted in this work, but they can set the absolute scale of the reported sensitivities. They should therefore be interpreted as defining a reference configuration, which can be refined or adapted for specific analyses. We describe these choices in more detail below.

\paragraph{Rejection threshold in the likelihood analysis} The rejection threshold (Sec.~\ref{sec:analysis}) is defined using the 95th percentile of the likelihood-ratio distribution for the null hypothesis (Fig.~\ref{fig:R0vsR}). This choice is primarily motivated by the need to define a measurable detection efficiency. Adopting a more stringent threshold (i.e., moving from $\sim 2\sigma$ to $5 \sigma$) would shift the inferred sensitivity toward larger gap amplitudes or higher subhalo masses, while a more permissive threshold would have the opposite effect. In practice, raising the threshold is also limited by the computational cost required to accurately sample the tail of the likelihood-ratio distribution (see Fig.~\ref{fig:R0vsR}). While the likelihood-ratio distribution in the idealized case follows the expected analytical behavior, the distribution must be empirically calibrated with simulations for the more realistic configurations. The $P(k)$ likelihood further neglects correlations between Fourier modes. 

\paragraph{Matched filter parameters}
The matched-filter parameters (see Sec.~\ref{sec:colormagcuts}) are fixed to follow true stream isochrone, corresponding to an idealized case. We verified that moderate variations of these parameters, which modify the selected background population and can change the relative contribution of contaminants such as galaxies, lead only to small shifts in the results.

\paragraph{Survey depth mitigation strategy}
We highlighted the need to mitigate survey depth variations, and we used the simplest approach of a bright magnitude cut.
To define this cut, we used the 2nd percentile of the local magnitude limits (Sec.~\ref{subsubsec:birghtmagcut}). While this choice is somewhat arbitrary, we tested several percentile thresholds and found the 2\% choice to provide a reasonable compromise for the configurations explored here.

\subsection{Minimal detectable subhalo masses in the literature}
\label{sec:litcomparison}
We presented estimates of the minimum detectable subhalo mass for \AAU-like streams using only linear density fluctuations along the stream. More generally, additional constraints can be obtained from stream kinematics, in particular velocity dispersion measurements \citep{nibauer:2025, carlberg:2025}, which have already been used to probe the subhalo population.

In \cite{lu:2025}, the minimum detectable mass is defined as the value for which $50\%$ of realizations are detected, which is equivalent to setting a criterion $P^{\mathrm{detect}}(M_{\mathrm{sub}})=0.5$ (Eq.~\ref{eq:proba_subhalM}). Their analysis additionally includes kinematic and morphological observables (e.g., proper motions, radial velocities, and stream track perturbations) but does not account for realistic survey selection, systematics or background contamination. Restricting the comparison to our ideal survey configuration (Table~\ref{tab:configurations}), we obtain $M^\mathrm{min}_\mathrm{sub} \simeq 5.3\times10^5\,M_\odot$ at $P^{\mathrm{detect}}\simeq0.5$, in good agreement with their \LSST-like results ($7.9\times10^5$--$1.6\times10^6\,M_\odot$). The remaining difference is consistent with our broader sampling of impact times, which includes older encounters producing stronger gaps, whereas their analysis fixes the impact to a recent epoch.

A complementary comparison can be made with Sec.~3.1.2 of \cite{drlica-wagner:2019}, who estimate the minimum detectable subhalo mass from a fixed $5\sigma$ signal-to-noise threshold after 10 years of \LSST observations, including stellar foregrounds but not mis-classified galaxy contamination. Their setup is therefore most comparable to our uniform survey configuration with perfect star-galaxy classification. For a surface brightness of $33\,\mathrm{mag/arcsec^2}$, they obtain $M_{\rm sub}^{\rm min} \simeq 2\times10^7\,M_\odot$, while we find $M_{\rm sub}^{\rm min} \simeq 4\times10^6\,M_\odot$ at $P^{\mathrm{detect}}\simeq0.5$. The remaining difference is consistent with their more conservative detection threshold and different survey assumptions.

\subsection{Stellar stream and background modeling}
\label{sec:limitations}

As motivated in Sec.~\ref{sec:streammodeling}, the stream, gap, and background were modeled in the configuration most favorable to detection; the reported detection efficiencies should therefore be interpreted as upper bounds, and the corresponding minimum detectable gap depths and subhalo masses as lower bounds. The relative degradation between survey configurations, the central result of this work, is comparatively more robust to these choices, since it is driven by systematics common to every configuration tested.

\paragraph{Other streams}
The properties of the simulated stellar streams have been fixed to those of \AAU. Since gap detectability primarily scales with the number of detected stream stars, these sensitivity estimates apply to streams with comparable observational properties, and would shift toward lower sensitivities for fainter or more distant systems such as Jet \citep{Jethwa:2018, ferguson:2021}. Variations in stellar population properties (e.g., age, metallicity, or isochrone) would also modify the selected background contamination, and therefore the absolute sensitivity values.

\paragraph{Stream parametric model}
The stream baseline is modeled as a uniform density profile along $\phi_1$ (Sec.~\ref{sec:stream_isochrones}), whereas real streams may exhibit more complex profiles arising from their dynamical history or a distance gradient along the stream. To quantify this, streams with a linear density gradient of $40\%$ along their length, consistent with the density variation expected from the distance gradient of \AAU \citep{Li:2021}, are injected and analyzed. The resulting minimum detectable gap depth differs by only $\sim1\%$ relative to the uniform baseline, confirming that the conclusions of this work are robust to deviations at first order from a uniform stream profile.

\paragraph{Uniform background}
The foreground star and background galaxy populations are modeled as spatially uniform (Sec.~\ref{sec:bkdg}), whereas foreground stars follow a Galactic density law with strong spatial variations, and background galaxies cluster in a way that can introduce additional spurious fluctuations \citep{tsiane:2025}. This assumption keeps the null hypothesis (Sec.~\ref{sec:analysis}) as simple as possible, requiring only a single nuisance parameter for the background level. As shown in the previous paragraph, a linear density gradient in the stream has a negligible impact on the detection efficiency; a comparable gradient in the background, expected to be even smaller at the high Galactic latitudes studied here, is therefore not anticipated to change our conclusions. Future work could relax this by sampling a more realistic Galactic background, for instance using the Besan\c{c}on model \citep{robin:2003}.

\paragraph{Gap model and subhalo-to-gap mapping}
Firstly, the gap is modeled with a simple box-car profile (Sec.~\ref{sec:gap-model}), which does not capture the full diversity of perturbations expected from subhalo-stream interactions \citep{erkal:2015}.  To go beyond this approximation, we repeated the analysis with a Gaussian gap profile which increases the minimum detectable gap depth by $\sim5\%$ (compared to the $\sim 3\%$ statistical errors, see Sec.~\ref{sec:results:sub:nobackground}), confirming the weak sensitivity to the gap morphology. Secondly, subhalo encounters are expected to redistribute stars across the stream in $\phi_2$ and to produce projection effects along the line of sight, neither of which is captured by the one-dimensional density framework adopted here. Finally, the subhalo study further relies on an analytical mapping between subhalo properties and gap observables \citep{erkal:2016}. While physically motivated, the latter remains an approximation compared to full dynamical modeling: it neither includes an explicit subhalo mass function, focusing on detectability rather than abundance, nor accounts for the possible tidal mass-dependent disruption of the subhalos \citep{2017PhRvD..95f3003S, 2019MNRAS.487.4409K, 2020MNRAS.499..116W, 2026PDU....5202262P}.

\subsection{Extensions and future work}
\label{sec:extensions}

\paragraph{Possible improvements and extensions}

We quantified the impact of realistic morphology-based star-galaxy classification implemented in the \LSST Science Pipelines. The significant loss of sensitivity induced by galaxy contamination highlights the importance of improving classification algorithms for stellar stream analyses. In this context, recent explorations into improved star-galaxy classification methods for LSST \citep[e.g.,][]{duan:2026, gatto:2026a} have the potential to improve the performance of our realistic configurations.

The simple magnitude cut used to mitigate the impact of survey depth variations is not fully optimized and does not aim to maximize the signal-to-noise ratio. 
More optimal mitigation approaches may alleviate part of this loss. In particular, methods that exploit the full set of survey property maps to correct for spatial inhomogeneities provide a promising alternative to hard magnitude cuts \citep{boone:2026}, potentially preserving a larger fraction of the available signal while controlling systematics.

The efficiency maps derived in this work, ${\cal E}(A,w)$ (i.e., Fig.~\ref{fig:efficiency_nobackground} and \ref{fig:efficiency_wbackground}),  provide a direct way to connect intrinsic density fluctuations to observable signatures.
They can be interpreted as a transfer function that maps a population of gaps, characterized by their depth and width, to the fraction that would be detectable in a \LSST-like survey.
In practice, these maps can be used to forward-model the observable gap distribution from a given theoretical prediction, by weighting each gap by its corresponding detection efficiency.
As a result, they can translate directly to streams with properties similar to those considered here (see Sec.~\ref{sec:streammodeling}), in particular in terms of distance and stellar mass (or, equivalently, number of observed stars) given in Sec.~\ref{sec:injected_parameters}. Since the sensitivity scales primarily with the number of stars, the efficiency maps are most reliable when used within this regime. For streams with significantly different properties, a dedicated survey forward modeling is required.

\paragraph{Towards realistic forecasts with dynamical simulations}

All the above limitations can be addressed in future work by combining this framework with fully computationally efficient dynamical stream simulations \citep[e.g.,][]{nibauer:2024}, more realistic background models, and specific DM scenarios. The methodology developed here is designed to accommodate such extensions, providing a flexible interface between theoretical predictions and observational constraints. 

Beyond these modeling limitations, the analysis strategy itself could also evolve depending on the scientific objective. In particular, the present framework is optimized for the detection of individual gap-like features in the linear density profile. Alternative summary statistics or metrics may become more appropriate for future analyses targeting the cumulative imprint of multiple perturbations, stacked stream populations, or joint constraints combining density and kinematic observables \citep[e.g.,][]{erkal:2015a,nguyen:2025, nibauer:2025}.

We stress that for more general applications, the most robust approach is to incorporate survey effects directly at the catalog level. This can be achieved by injecting dynamical stream simulations into a realistic survey framework, and applying the full sequence of observational effects: photometric uncertainties, detection and classification efficiencies, background contamination, and mitigation strategies such as magnitude cuts.
The \StreamObs package developed in this work is designed to perform this task (Sec.~\ref{sec:genobs}), providing a consistent way to transform intrinsic simulations into \LSST-like observations. 

\section{Conclusion}
\label{sec:conclusion}

In this work, we developed a forward-modeling framework to quantify the impact of \LSST observational effects on the detection of density fluctuations in stellar streams and on the corresponding sensitivity to DM subhalo encounters. Implemented within the public \StreamObs package, this framework provides a consistent way to propagate intrinsic stream models into realistic \LSST-like observations, including photometric uncertainties, selection effects, survey depth variations, and background contamination.

We find that observational effects have a major impact on stream gap detectability. For a representative stream with properties similar to \AAU, background contamination is the dominant limitation, degrading the sensitivity to density fluctuations by a factor of $\sim4$ compared to an idealized background-free case. Survey systematics, including depth variations and imperfect star-galaxy classification, further amplify this loss in sensitivity by $\sim 23\%$. Together, these effects degrade the minimum detectable subhalo mass by a factor of up to $\sim16$ after $4$ years of \LSST observations, with star-galaxy misclassification as the dominant systematic (factor of $\sim2.5$), leading to a lightest detectable subhalo of $\sim1\times10^7$\,M$_\odot$.
This shows that forecasts based solely on dynamical stream simulations---with stars selected down to the nominal survey depth---are too optimistic.
For a similar survey model, our sensitivity estimates remain consistent with previous forecasts (Sec.~\ref{sec:litcomparison}), confirming that this framework reproduces existing results while explicitly quantifying the additional impact of realistic survey effects.

We demonstrated that mitigation strategies are necessary to suppress survey-induced inhomogeneities.
They are also required to achieve a sensitivity gain when moving from 1-year to 4 years of \LSST observations. The simplest mitigation strategy is to apply magnitude cuts, which trade statistical power for control of survey systematics.
As a step towards performing a more realistic analysis, we explore the use of such cuts on simulations. 
Namely, these cuts correspond to restricting stellar stream analyses to objects with $g < 24.2\magn$ ($r < 24.6\magn$) for 1 year of observations and $g < 25.3\magn$ ($r < 25.6\magn$) for 4 years of observations. Such cuts provide a more realistic approximation than simply selecting stars down to the nominal $5\sigma$ survey depth, although they do not capture other observational limitations such as background contamination or imperfect star-galaxy classification.

Better mitigation strategies will be needed to fully benefit from the unprecedented depth of \LSST. 
Recently, \citet{boone:2026} proposed a technique to use synthetic-source injection to correct for survey non-uniformity, though such a procedure comes with significant computational cost.
On the other hand, survey uniformity will improve naturally with the later \LSST data releases \citep[e.g.,][]{awan:2016,leistedt:2026}.
Additional progress is also expected from improved star-galaxy classification, which currently limits the usable depth of the survey at faint magnitudes. Future synergies with higher-spatial-resolution surveys (e.g., Euclid and Roman) and/or complementary multi-band observations, could significantly improve the classification performances and recover part of the currently inaccessible \LSST statistical power \citep{han:2023}.
Finally, the sensitivity of \LSST stellar stream observations to DM subhalos could also be enhanced by combining density fluctuation measurements with additional observables such as proper motions and stream track perturbations \citep[e.g.,][]{erkal:2015a,nguyen:2025}; exploiting such astrometric information is left to future analyses, as it goes beyond the photometric framework developed here.

In this study, we include the effects of survey depth non-uniformity, star-galaxy misclassification, and background contamination to generate realistic mock observations of stellar streams with \LSST.
We considered the impact of these survey systematics on a simple observable, namely the detectability of gaps in the density of stellar streams.
Our results show that background contamination and survey systematics strongly affect both gap detectability and subhalo mass sensitivity, implying that forecasts neglecting these effects can lead to overly optimistic conclusions.
Our analysis framework could be improved to include more complex metrics, such as measurements of the two-dimensional density power spectrum which captures the ensemble of DM subhalo perturbations.
Furthermore, our mock data generation framework can be applied to dynamical streams generated in a Milky-Way-like potential \citep[e.g.,][]{shipp:2023, pearson:2024,  arora:2026a}.
Finally, the framework presented here could be naturally extended to combine additional observational modalities that will be available from photometric, astrometric, and spectroscopic surveys in the coming decade.
To conclude, we emphasize that it is essential to realistically model observational effects to rigorously assess the sensitivity of future stellar stream observations to DM subhalo properties.

\textit{Software:} \healpix \citep{Gorski:2005}\footnote{\url{http://healpix.sourceforge.net}}, \code{healpy}\footnote{\url{https://github.com/healpy/healpy}}, \code{iMinuit} \citep{dembinski_2024_13902219}, \numpy \citep{harris2020array}, \code{pandas} \citep{mckinney-proc-scipy-2010,the_pandas_development_team_2025_17229934}, \code{RubinSim}, \code{Scipy} \citep{2020SciPy-NMeth},  \code{skyproj}\footnote{\url{https://github.com/LSSTDESC/skyproj}}, \code{ugali} \citep{Bechtol:2015}.

\section*{Acknowledgments}


This paper has undergone internal review in the LSST Dark Energy Science Collaboration. 
The internal reviewers were Christian Aganze and Johann Cohen-Tanugi.

Author contributions were as follows: M.P. led the analysis, software development, figure generation, and manuscript writing. P.S.F. contributed to software development, figure construction, and manuscript writing. A.D.W. conceptualized and supervised the project, contributing to software development, figure review, and manuscript review. M.K. conceptualized and supervised the project, contributing to software development/review, figure and manuscript review. D.M. conceptualized and supervised the project, contributing to figure and manuscript review. C.A. and J.C.T. contributed to manuscript review. Y.-Y.M. is a DESC builder and contributed to the generation of the DC2 truth-match catalog used in this work.


PSF acknowledges support from the DiRAC Institute in the Department of Astronomy at the University of Washington.
The DiRAC Institute is supported through generous gifts from the Charles and Lisa Simonyi Fund for Arts and Sciences, Janet and Lloyd Frink, and the Washington Research Foundation.

The DESC acknowledges ongoing support from the Institut National de 
Physique Nucl\'eaire et de Physique des Particules in France; the 
Science \& Technology Facilities Council in the United Kingdom; and the
Department of Energy and the LSST Discovery Alliance
in the United States.  DESC uses resources of the IN2P3 
Computing Center (CC-IN2P3--Lyon/Villeurbanne - France) funded by the 
Centre National de la Recherche Scientifique; the National Energy 
Research Scientific Computing Center, a DOE Office of Science User 
Facility supported by the Office of Science of the U.S.\ Department of
Energy under Contract No.\ DE-AC02-05CH11231; STFC DiRAC HPC Facilities, 
funded by UK BEIS National E-infrastructure capital grants; and the UK 
particle physics grid, supported by the GridPP Collaboration.  This 
work was performed in part under DOE Contract DE-AC02-76SF00515.

\bibliographystyle{aasjournal}
\bibliography{references_lsst_syst}

\appendix

\section{Analysis comparison in real and Fourier space}
\label{app:poisson}
We perform the main gap detection analysis in Fourier space using the power spectrum likelihood (Sec.~\ref{sec:analysis}), and present here the corresponding real-space results for comparison. This complementary approach serves as a validation test, ensuring that the results obtained with the Fourier-space likelihood are consistent with those derived from a more standard and widely used Poisson framework.

\paragraph{Likelihood in real space}

The analysis in real space uses a Poisson likelihood defined as
\begin{equation}
    \mathcal{L}_{\mathrm{poisson}} = \prod_{i} 
    \frac{(N_i^{m})^{N_i} \, e^{-N_i^{m}}}{N_i!} \, ,
    \label{eq:likelihood_poisson}
\end{equation}
where $N_i$ is the observed number of objects and $N_i^{m}$ is the expected number of objects predicted by the model in bin $i$.

\paragraph{Impact of likelihood choice on efficiency maps}
\begin{figure*}[t]
\centering
\includegraphics[width=\textwidth]{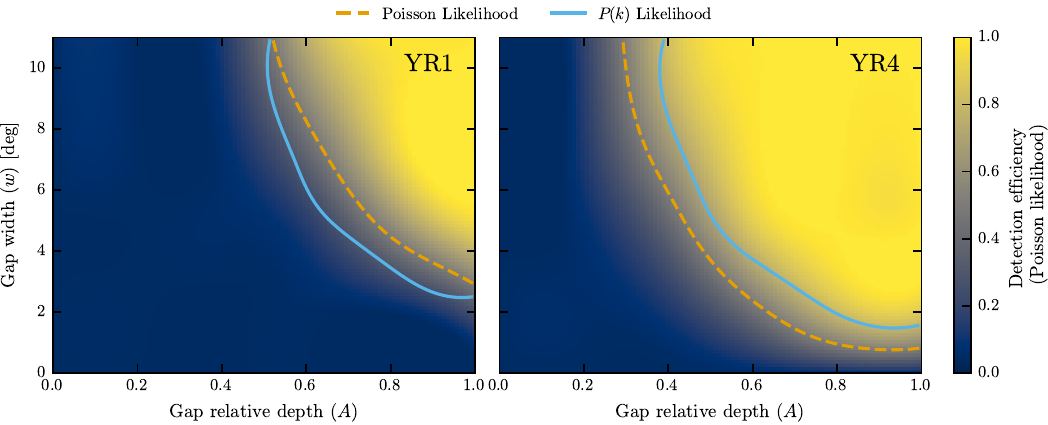}
\caption{Same as Fig.~\ref{fig:efficiency_wbackground}. It shows the gap sensitivity detection in various scenarios, in the presence of background contamination, calculated from the analysis in Fourier (solid blue line) and real (dashed orange line) space, for the most realistic survey configuration only (inhomogeneous survey, background and S/G contamination).}
\label{fig:efficiency_wbackground_poisson}
\end{figure*}

We compare in Fig.~\ref{fig:efficiency_wbackground_poisson} the impact of moving the analysis from $P(k)$ to the Poisson likelihood; Eqs.~\eqref{eq:likelihood} and~\eqref{eq:likelihood_poisson} respectively. For readability, we show the most realistic configuration only.
The $50\%$ efficiency iso-contours are broadly consistent with one another. We have also checked (not shown) that the main conclusions derived from the power-spectrum likelihood also hold for a standard real-space analysis.

This validates the Fourier-space framework developed in this work, which shows performances similar to that of the traditional Poisson likelihood analysis.
The differences observed can arise from the approximation made in the Fourier-space likelihood, which neglects correlations between modes. In addition, the two approaches probe the data differently: the real-space Poisson likelihood is mainly driven by the significance of localized individual gaps, whereas the Fourier analysis is sensitive to the overall fluctuation power across multiple scales, including fluctuations induced by survey systematics. As a result, the Poisson approach can be more sensitive when a single prominent gap dominates the signal, while the Fourier analysis provides a more global characterization of the stream fluctuations.

\paragraph{Impact of likelihood choice on gap depth sensitivity}

\begin{figure*}[t]
\centering
\includegraphics[width=0.45\textwidth]{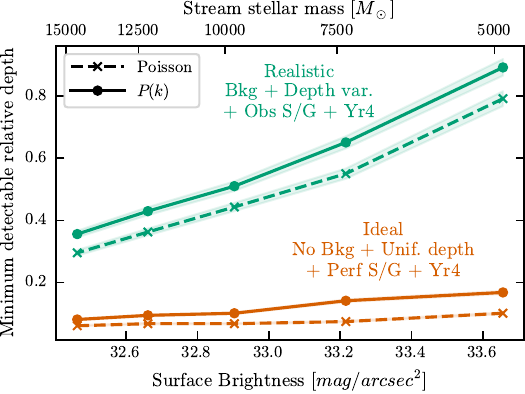}
\includegraphics[width=0.45\textwidth]{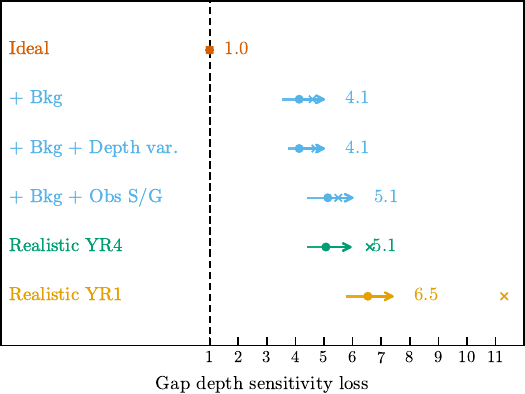}
\caption{Similar to Fig.~\ref{fig:gap_sensitivity}, with results obtained using the Poisson likelihood \eqref{eq:likelihood_poisson} overplotted. Left shows the minimal detectable relative amplitude, obtained from the efficiency condition ${\cal E}(A^{\rm min},w=5\deg)=0.5$, as a function of stream stellar mass (top axis) and surface brightness (bottom axis) after 4 years of observation. For readability, we only show the result for the realistic (green lines) and ideal (orange lines) survey configurations, for both the $P(k)$ (solid lines) and Poisson (dashed lines) likelihoods. Right panel presents the ratio $A^{\mathrm{min}} / A^{\mathrm{min}}_{\mathrm{ideal}}$, highlighting the degradation induced by different survey systematics. Markers indicate the reference case with $N^{\mathrm{true}}_\star = 4 \times 10^{4}$ stars, while error bars and arrows reflect the dependence on surface brightness.}
\label{fig:gap_sensitivity_alsoPoisson}
\end{figure*}

We show in the left panel of Fig.~\ref{fig:gap_sensitivity_alsoPoisson}, for the two likelihood choices, the  minimal detectable gap relative depth (from  ${\cal E}(A_{\mathrm{min}},\,N^{\rm true}_\star,\,w\!=\!5^{\degree})=0.5$) as a function of the stream surface brightness. 
Both likelihood analyses exhibit similar trends. The use of the Poisson likelihood slightly decreases the minimal detectable gap depth for the realistic and ideal surveys. 

The right panel summarizes our results, in terms of gap depth sensitivity loss. When the Poisson likelihood is used, these numbers are slightly larger (crosses compared to disks). We therefore adopt the power spectrum analysis results throughout the main text, as they provide a more conservative estimate of the sensitivity loss.

\section{Depth based magnitude cuts}
\label{app:magcut}

\begin{figure*}[t]
\centering
\includegraphics[width=0.45\textwidth]{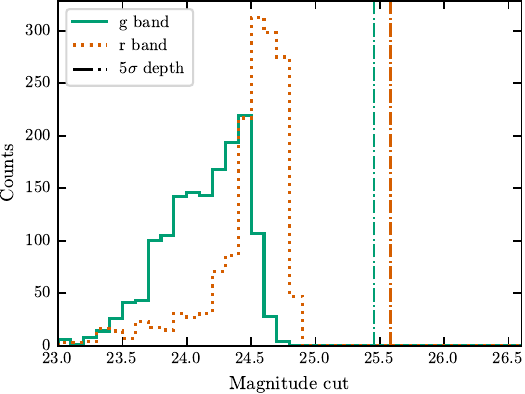}
\includegraphics[width=0.45\textwidth]{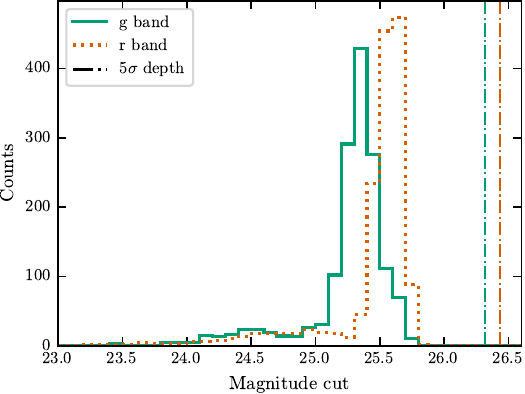}
\caption{Distribution of the adaptive magnitude cuts used to mitigate survey depth variations, estimated from the local magnitude limits around injected uniform \AAU-like streams without background contamination. Results are shown for the $g$ and $r$ bands after 1 year (left) and 4 years (right) of \LSST observations. The vertical dashed lines indicate the median $5\sigma$ magnitude limits in each band.}
\label{fig:magcut}
\end{figure*}

Figure~\ref{fig:magcut} illustrates the distribution of the adaptive magnitude cuts described in Sec.~\ref{subsubsec:birghtmagcut}. In both survey releases, the cuts are brighter in the $g$ band than in the $r$ band, reflecting the shallower depth of the survey in $g$. The adopted cuts also remain brighter than the median $5\sigma$ magnitude limits, and are typically slightly more restrictive than the corresponding $10\sigma$ depths (obtained by shifting $5\sigma$ magnitude limits by $-0.75\magn$). The distributions shift toward fainter magnitudes between Y1 and Y4, consistent with the increased survey depth after additional years of observations.

\section{Subhalo mass to gap properties}
\label{app:subhalosproba}

The mapping between subhalo properties and gap observables (amplitude $A$ and width $w$) is computed using the analytical framework of \cite{erkal:2016}. For completeness, we reproduce below the relevant formulae. We also give the sampling strategy used to calculate the conditional probability $P(A,w \mid M_{\mathrm{sub}})$, used in Sec.~\ref{sec:subhalomass}. 

After an interaction time $t$, the gap depth is given by
\begin{equation}
A = 1 - 
\left(1 + \frac{4-\gamma^2}{\gamma^2}
\frac{v_\perp^2}{v^3}
\frac{2GM_{\mathrm{sub}}}{b^2 + r_s^2}
t \right)^{-1}\,,
\label{eq:gapdeth}
\end{equation}
where $b$ is the impact parameter, $v_\perp$ the velocity perpendicular to the stream,  $v = (v_\parallel^2 + v_\perp^2)^{1/2}$ the total relative velocity, 
$r_s =  \sqrt{M_{\mathrm{sub}}/10^8\,M_\odot} \times 1.62\,\mathrm{kpc}$ the scale radius of the impacting subhalos (modeled as Plummer spheres) of mass $M_{\mathrm{sub}}$, and $\gamma^2 = 2$ for a locally flat rotation curve of the Milky-Way potential.
The gap width evolves through two regimes separated by the caustic time
\begin{equation}
    t_{\mathrm{caustic}} =
    \frac{4\gamma^2}{4-\gamma^2}
    \frac{v^3}{v_\perp^2}
    \frac{b^2 + r_s^2}{G M_{\mathrm{sub}}} \, .
\end{equation}
Before this transition, the width grows as
\begin{equation}
    w =
    2\frac{v}{v_\perp}\frac{\sqrt{r_s^2 + b^2}}{r_0}
    + \frac{2GM_{\mathrm{sub}} v_\perp}{v^2 r_0 \sqrt{r_s^2 + b^2}}
    \frac{4-\gamma^2}{\gamma^2} t \, ,
\end{equation}
\\
while at later times it follows
\begin{equation}
    w =
    4 \left(
    \frac{4-\gamma^2}{\gamma^2}
    \frac{2GM_{\mathrm{sub}}}{v r_0^2} t
    \right)^{1/2} .
\end{equation}

\paragraph{Sampling of gap properties}
For a given subhalo mass, a distribution of gap properties $(A,w)$ is obtained by sampling the interaction parameters. To probe the expected impact for our baseline stellar stream \AAU we use the observed midpoint of the stream to define the 6D phase-space position of \AAU \citep{Li:2021, Mateu:2023}. 
Following \citet{erkal:2016}, we make the simplifying assumption that the stream is on a circular orbit finding the Galactocentric radius is the time averaged radius of AAU equal to $r_0=28.5\,\kpc$. 
Then the stream velocity is just the tangential velocity of the stream $v_s = 174\,\mathrm{km\,s^{-1}}$.
Finally, the stream age, $t_s = 5\,\mathrm{Gyr}$, is the dynamical age needed to generate the stream in \citet{Li:2021}.

The impact time is drawn uniformly between $0$ and the dynamical age of the stream $t_s$.
The velocity components are sampled following \cite{erkal:2016}, assuming a subhalo velocity dispersion $\sigma_{\mathrm{sub}} = 180\,\mathrm{km\,s^{-1}}$.
The impact parameter is fixed to $b = 0$, providing a conservative estimate of detectability by maximizing the induced perturbation. 
The impulse approximation is enforced by requiring \citep{erkal:2015}
\begin{equation}
    \frac{v}{v_\perp} \frac{\sqrt{b^2 + r_s^2}}{r_0} \ll 1 \, .
\end{equation}
\\
The sampling is restricted to the range of gap sizes explored in the efficiency analysis. In practice, configurations with $w > 11\degree$ are negligible and are excluded.
This procedure defines the conditional distribution $P(A,w \mid M_{\mathrm{sub}})$, which describes the range of gap properties produced by a subhalo of mass $M_{\mathrm{sub}}$. 
It does not encode the abundance of subhalos, but only the mapping between mass and observable signatures. 

\end{document}